\documentclass[9pt,twocolumn,twoside]{IEEEtran}
\usepackage[ruled,vlined]{algorithm2e}
\usepackage{amssymb}

\usepackage{amsmath}
\usepackage{array}
\newcolumntype{C}[1]{>{\centering\arraybackslash}p{#1}}

\DeclareSymbolFont{letters}{OML}{ztmcm}{m}{it}
\DeclareSymbolFontAlphabet{\mathnormal}{letters}

\usepackage{balance}
\usepackage{boldline}			
\usepackage{color}
\usepackage[inline]{enumitem}
\usepackage{balance}
\usepackage[rawfloats=true]{floatrow}
\usepackage[symbol]{footmisc}

\usepackage{graphicx}
\usepackage{listings}
\usepackage{multibib}
\usepackage{multicol}
\usepackage{multirow}
\usepackage{nicefrac}
\usepackage{placeins}
\usepackage{rotating}
\usepackage{resizegather}
\usepackage[caption=false]{subfig}
\usepackage{textcomp}
\usepackage{times}
\usepackage{url} 
\usepackage{xspace}
\usepackage{tabularx}
\usepackage[flushleft]{threeparttable}
\usepackage{color}  
\usepackage{xcolor}
\usepackage{todonotes}
\usepackage{hyperref}
\hypersetup{
    colorlinks=true, 
    linktoc=all,     
    linkcolor=blue,  
    urlcolor=blue,
}

\usepackage{cite}

\usepackage{pifont}
\newcommand{\cmark}{\ding{51}}%
\newcolumntype{P}[1]{>{\centering\arraybackslash}p{#1}}

\newcommand{\Csharp}{%
  {\settoheight{\dimen0}{C}C\kern-.05em \resizebox{!}{\dimen0}{\raisebox{\depth}{\#}}}}
  
\graphicspath{{./graphics/}{./graphics/new_charts/}}

\usepackage{fancyhdr,lipsum}
\fancypagestyle{firstpage}{
  \fancyhf{}
  \fancyhead[C]{THIS IS AN EXTENDED VERSION OF THE ACCEPTED CONFERENCE PAPER}
}
\title{Caching Structures for Distributed Data Management in P2P-based Social Networks}

 \author{\IEEEauthorblockN{Newton Masinde, Moritz Kanzler and Kalman Graffi}
\IEEEauthorblockA{\\Technology of Social Networks\\Heinrich Heine University,\\ Universit{\"a}tsstrasse 1, 40225 D{\"u}sseldorf, Germany\\email: \{newton.masinde, moritz.kanzler, graffi\}@hhu.de\\web: http://tsn.hhu.de/}

}

\begin{document}

\maketitle \thispagestyle{firstpage}

\begin{abstract}
 Distributed applications require novel solutions to tackle problems that arise due to the scarcity of resources such as bandwidth, memory and processing power.
 One of these challenges is seen in distributed data management.
 The challenge is the two part problem of ensuring that the content is valid when accessed and updating it immediately when changed.
 This is especially difficult when considering p2p-based distributed online social networks, which aim to build reliable, secure social networking platforms on top of often unreliable and unsecure devices.
 In this paper, we propose three selection strategies, random, trend and social score, for a social caching mechanism. 
 They consider the social interaction patterns in the social network. 
 We implement and evaluate them in a DHT-based distributed online social networks called LibreSocial and show that the social score is the best strategy.
 Further we implement the social caching solution and also show that when used in combination with the existing caching solution almost all requests can be serviced via cache while retaining the consistency of data during updates.
\end{abstract}

\section{Introduction}
\label{sec:Introduction}
Online social networks have revolutionized how users communicate with each other and share content as well as information.
As a consequence, there has been a rapid growth in the amount of information that is available online, as well as the volume of content that users upload into the various social media platforms.
The obvious downside to this phenomenal increase in shared content and the connections initiated is seen in the growing concerns raised regarding the privacy of personal information and the ownership of uploaded content as well as the high costs of scaling the network.
These concerns are a direct result of the use of centrally managed systems by the online social network (OSN) system providers.
The system providers essentially own all the content uploaded to their servers, including private information and in many cases utilize this information to generate an income by selling it to third parties.
As a solution to these concerns, distributed online social networks (DOSNs) have been proposed.
A distributed online social network is designed with the goal of ensuring that the OSN operates in a distributed fashion with minimal or no central control.
This design offers three key benefits over the centralized OSNs:
\begin{enumerate*}[label=\itshape\alph*\upshape)]
\item
 general reduction on the operational costs as most resources are provided by the users; 
\item
 possibility for implementing better user-oriented privacy control; and 
\item
 innovative development~\cite{BuD09} as more resources, such as communication and storage, are availed via the users.
\end{enumerate*}

The gains realized in offering solution to mitigate the privacy and scaling concerns of centralized OSNs by the DOSNs are met by challenges in the actual design and deployment of the DOSNs.
One of these challenges is the management of social content or data in a distributed environment~\cite{GCP+18} which can also be viewed as the problem of \textit{scalable dissemination of social updates}~\cite{HPN+12}.
``Social data'' or ``social update'' is basically all the data that is exchanged among the users (profile information, relationships, community memberships and so on) and the generated content (comments, messages, posts, images and so on)~\cite{HPN+12,GCP+18}.
While a content delivery network (CDN) is used in a centralized network to help reduce the load on the network, the selection of a good social content delivery strategy can effectively achieve load reduction on the content storing nodes.
One method of handling the dissemination of this social data is the use of social replication/caching of the data.
Social caching differs from distributed caching~\cite{BGW10} and dynamic data replication~\cite{AZd93} on several points~\cite{HNI+11}: social network topology shows non-trivial clustering and positive degree correlation~\cite{NPa03}, social cache selection is social-relationship driven, and social caches only cache updates for friends to ensure a predictive pre-loading and one-hop communication.

In this paper, we propose a social caching mechanism for a P2P-based DOSN application called \textit{LibreSocial} (previously called LifeSocial.KOM~\cite{GPM+08,GMM+09,GGM+10,GGS11}) with the aim of reducing the load on the content storing nodes.
We propose three caching selection strategies called \textit{random algorithm}, \textit{trend algorithm} and \textit{social-score algorithm}, and compare the performance of each of these selection strategies.
The social-score algorithm is designed to utilize the social graph while taking into account the various centrality measures within a nodes ego network.
We show that the social-score selection strategy outperforms the others based on the cache hit ratio.
Further, we evaluate the overall performance of the network taking into account four instances, that is, no cache, default cache, social cache and social cache plus default cache.
The results show that the network gives the best values when the default and social cache are enabled together.

\section{Related work}
\label{sec:related_work}
There have been several ways in which DOSNs have been classified such as based on the decentralized data model~\cite{DBV10}, privacy and security considerations~\cite{GKB12,TKO15,DMR18}, and design decisions such as storage, access control and interaction~\cite{PFS14}.
These classifications, though comprehensive do not provide insightful information about how data management in terms of handling social data is performed in DOSNs.
Probably the best comprehensive study on social data management in DOSN is given in~\cite{GCP+18} in which DOSNs are classified into three groups, namely \textit{DHT-based}, \textit{social overlay (SO) based} and \textit{external resource-based} DOSNs.

\textit{DHT-based DOSNs}, such as DECENT~\cite{JNM+12}, Cachet~\cite{NJM+12}, and LibreSocial~\cite{GPM+08,GMM+09,GGM+10,GGS11},  are characterized by an overlay that relies on a distributed hash table (DHT), which is build through all participating peers and provides a key-value storage as well as ID-based routing. 
The DHT usually is used for storing the social content and also offers indexing services.
On the other hand, for \textit{social overlay (SO) based DOSNs}, the social overlay is a network of logical connection between pairs of nodes that correspond to their friendship relations, and they also exploit the DHT for indexing services.
An examples of a SO-based DOSN is DiDuSoNet~\cite{GAD+16}.
Finally, {external resource-based DOSNs} like Diaspora (\url{https://diasporafoundation.org}) and Vis-\'{a}-vis~\cite{SLC+11}, are set up through a federation of private web serves and thus are a compromise between centralized servers and completely decentralized solutions.

The appropriate solutions that help in ensuring reliable decentralized data management services for DOSNs are a great necessity that must always be kept in mind.
However, it requires handling two challenges associated with are \textit{data availability/persistence} and \textit{information diffusion}~\cite{GCP+18}.
In the context of this work , we discuss these two points in detail and give a summary on privacy.

\begin{figure}[!tbp]
  \centering
  \begin{minipage}[b]{0.99\linewidth}
        \centering    
        \includegraphics[height=2.9cm]{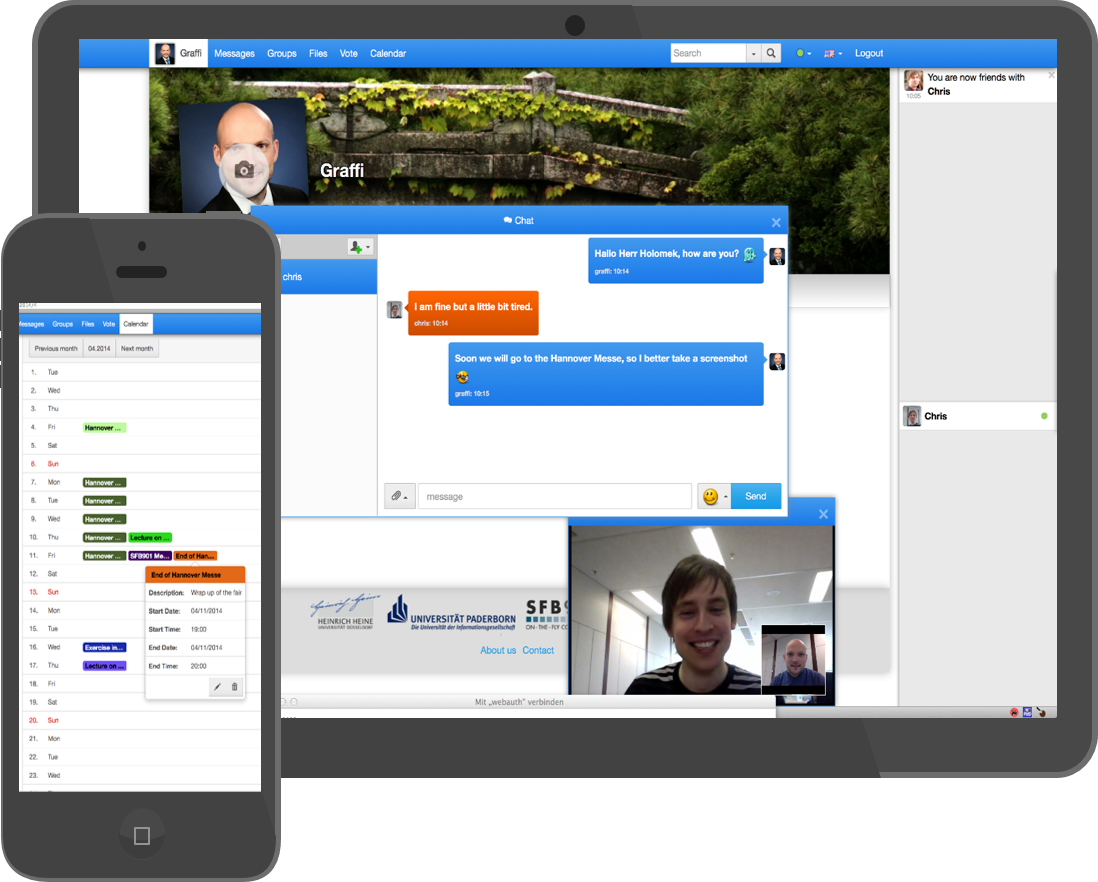}
        \includegraphics[height=2.9cm]{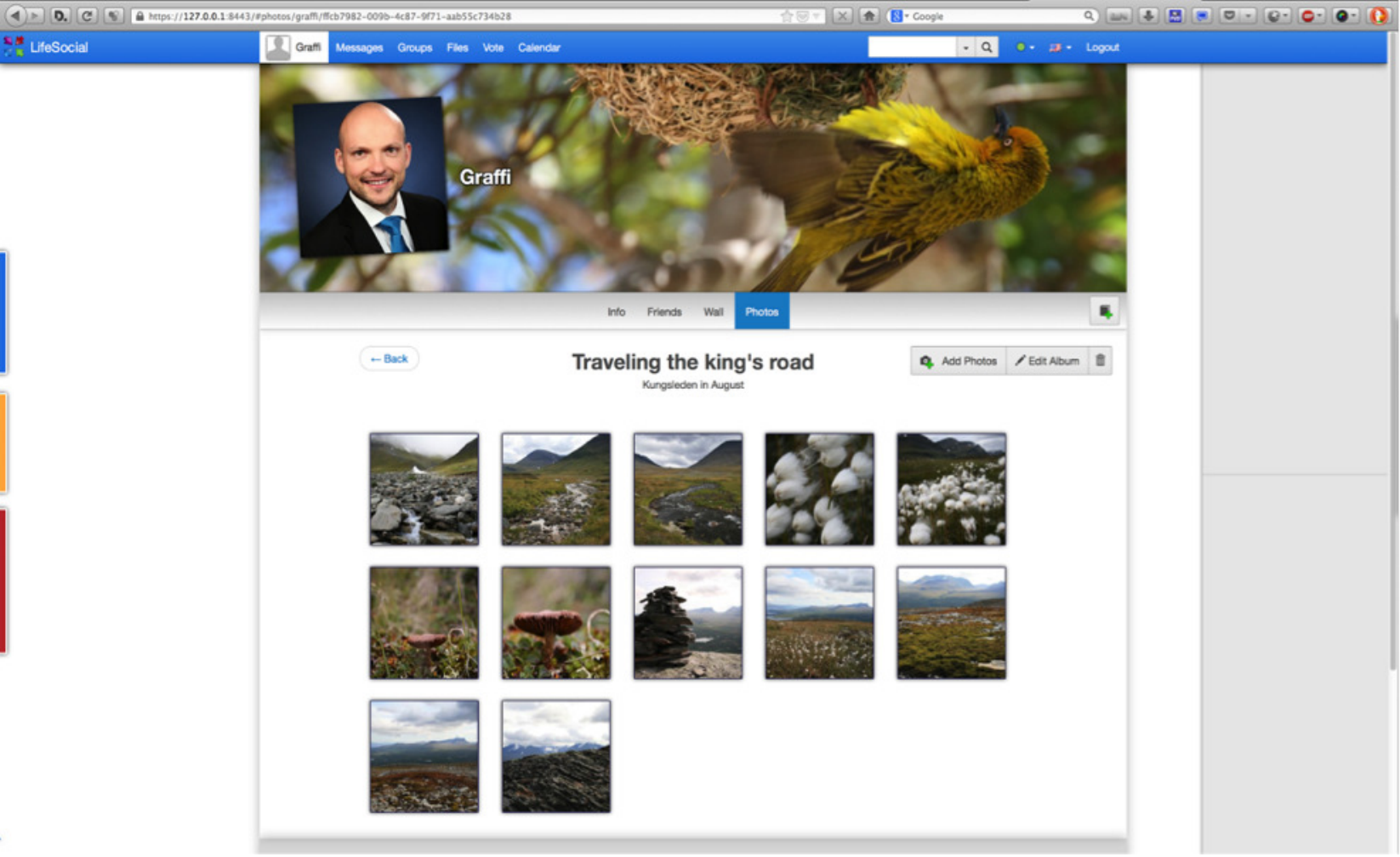}

        \includegraphics[height=2.9cm]{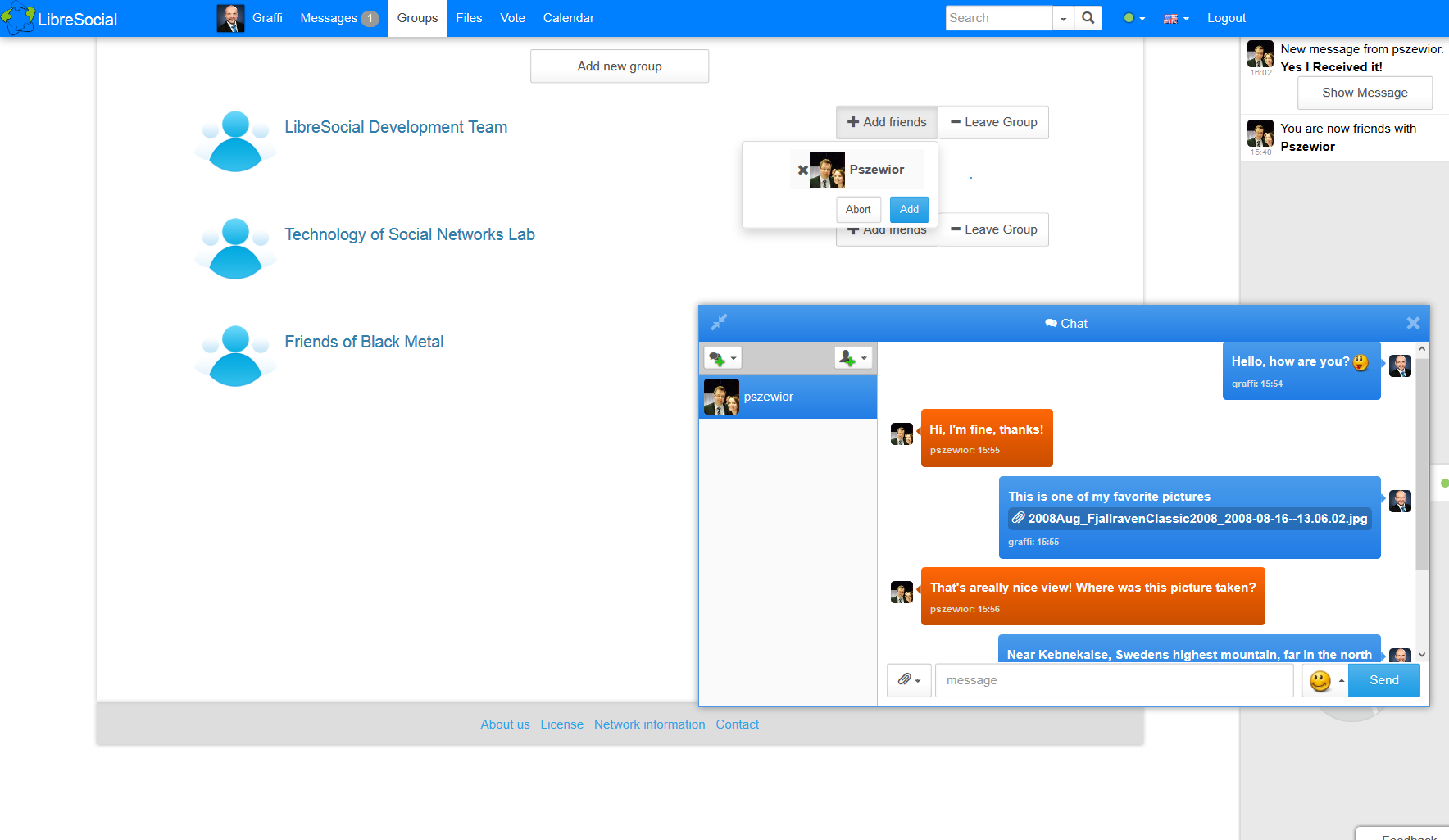}
        \includegraphics[height=2.9cm]{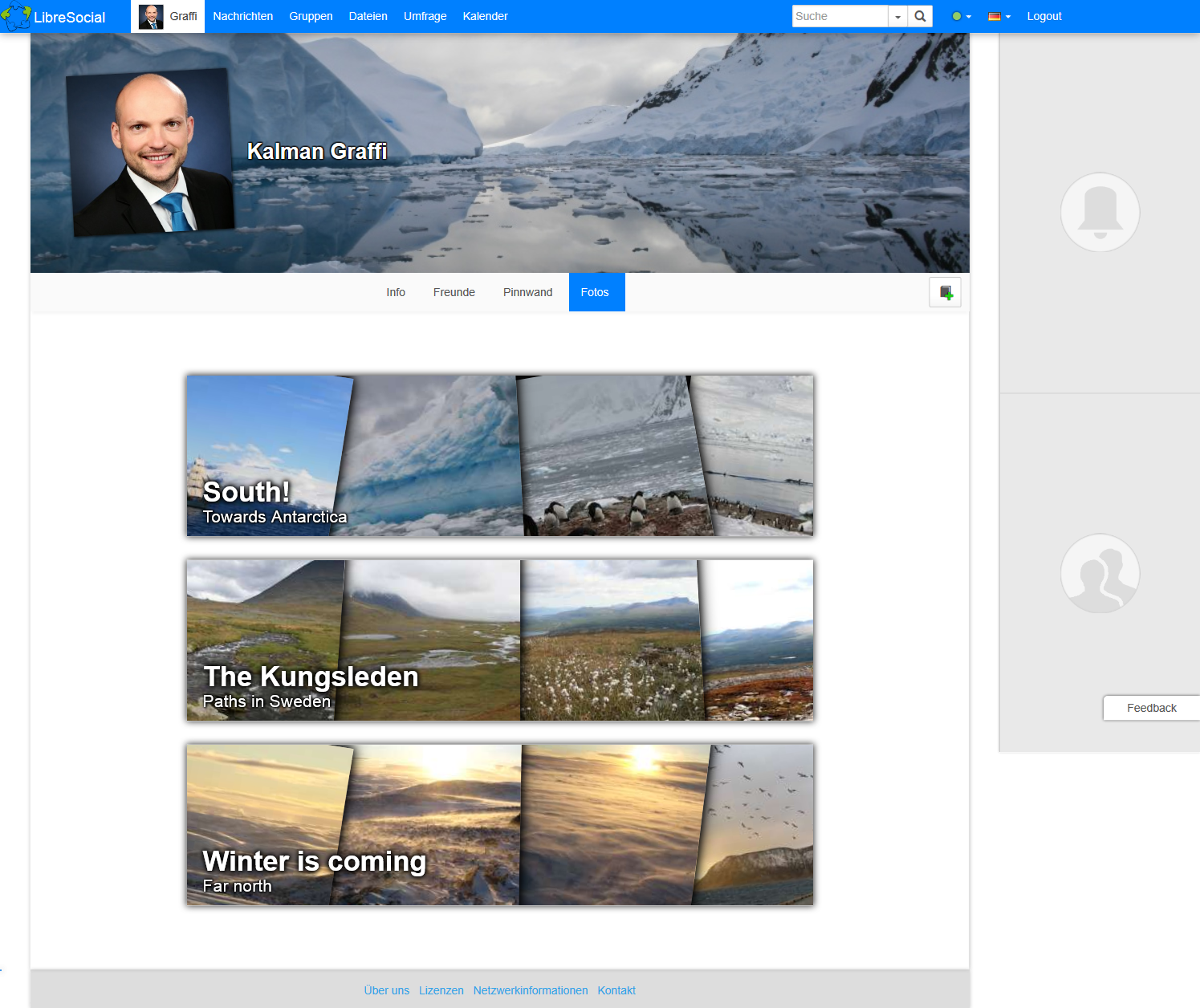}
        \caption{Screenshots of LibreSocial}
        \label{fig:LibreSocial_Screenshot}
  \end{minipage}
  \vspace{-0.4cm}
\end{figure}

\subsection{Data availability}
\label{subsec:DataAvailability}
This challenge entails the persistence of publicly accessible data in a distributed environment for which several techniques such as distributed dynamic data replication~\cite{AZd93}, erasure codes~\cite{CGD+14} and caching~\cite{BGW10} are proposed.
However these traditional approaches do not work well for DOSNs.
Thus, suitable solutions in DOSNs can be divided based on two considerations, \textit{external resource-based strategies} and \textit{replica selection strategies}~\cite{GCP+18}.

In external resource-based DOSNs, the data availability is guaranteed by utilizing some external resource management service such as cloud-based storage services, or even hybrid solutions that entail distributed storage augmented with cloud services.
Cadros~\cite{FHL+14} is an example of a DOSN that uses cloud assisted data replication.
Replication requires the design of a replica selection strategy to guarantee availability and the strategies can be separated into two general groups, \textit{replication in the DHT (untrusted nodes)} and \textit{replication on trusted nodes}.
In \textit{replication in the DHT (untrusted nodes)}, replication of stored content is handled by the underlying DHT overlay, therefore the social network has no control whatsoever of the replica placement.
In LibreSocial the content stored in the DHT is encrypted and a sophisticated access control scheme is applied, thus the data storage nodes do not have to be trusted. 
On the other hand, \textit{replication on trusted nodes} does not depend on the overlay replication mechanisms.
Replica placement is controlled by the social network, where each node selects a suitable nearby replica node based on some calculated measurements.
Two key aspects must be accounted for in this replication strategy.
The first aspect is that nodes can go offline at any time which can affect persistent storage.
This can be handled by analysis of online behavior of the nodes.
The second aspect is how to gauge the trust between nodes so that data can be exchanged.
This can be solved by inclusion of encryption as well as determination of social strengths that peers have with one another based on factors such as number of common friends or number of actual interactions between each other.

\subsection{Information diffusion}
\label{subsec:InfoDiffusion}
The second challenge to consider is information diffusion.
This covers the possibilities available in efficient access of needed information.
There, costs have to be taken into account in term of factors such as network bandwidth, response times and so on, while ensuring data consistency.
In the case of DOSNs, this challenge relates to the update dissemination process wherein updates to a user's data are made available to all interested nodes~\cite{GCP+18}. 
The update dissemination process can fall into three classes, \textit{request-reply}, \textit{active dissemination} and \textit{hybrid} approaches

\textit{Request-reply approaches} are based on the notion that, if data availability is guaranteed, all available content can be accessed by simply making a request for it.
The node responsible for the content will then reply by availing the requested content to the requester and in case the owner of the content is offline, replica nodes will give a response.
This approach is seen in DOSNs such as PeerSoN, DECENT, SocialCDN~\cite{HPN+12}, Vegas and LibreSocial.
\textit{Active dissemination approaches} may be viewed as a form of network flooding common in unstructured P2P networks but the distribution of the updates is limited using mechanisms such as gossip protocols like rumor mongering and anti-entropy~\cite{MMP11}, or based measures such as weighted ego betweenness centrality (WEBC)~\cite{CDG+14}, social interaction behavior~\cite{TVa12} and so on, to ascertain that only the nodes that require the update receive it.
Examples of DOSNs that use this approach include DiDuSoNet.
The last class, \textit{hybrid approaches}, the DOSNs implement a combination of request-reply and active dissemination approaches.
This is seen in Cachet~\cite{NJM+12}. 


\section{Data management in LibreSocial}
\label{sec:LibreSocialDataMgmt}
LibreSocial, previously LifeSocial.KOM\cite{GPM+08,GMM+09,GGM+10,GGS11}, is a DOSN designed based on a P2P framework model, hence relies on a DHT for its underlying functionality.
Screenshots of LibreSocial are presented in \ref{fig:LibreSocial_Screenshot}. 
LibreSocial is designed using (a heavily modified) FreePastry (\url{http://www.freepastry.org/FreePastry}), an open source implementation of Pastry~\cite{RoD01} which includes PAST~\cite{DRo01}, a persistent storage utility that manages data replication.
PAST relies on FreePastry for routing, and hence LibreSocial performs replication within the structure of the DHT. LibreSocial's architecture has been previously discussed.
We focus here primarily on how LibreSocial achieves data management, and specifically its storage structure.

\subsection{Storage structure of LibreSocial}

\begin{figure}[!tbp]
  \centering
  \begin{minipage}[b]{0.47\linewidth}
        \centering
        \includegraphics[width=0.99\linewidth]{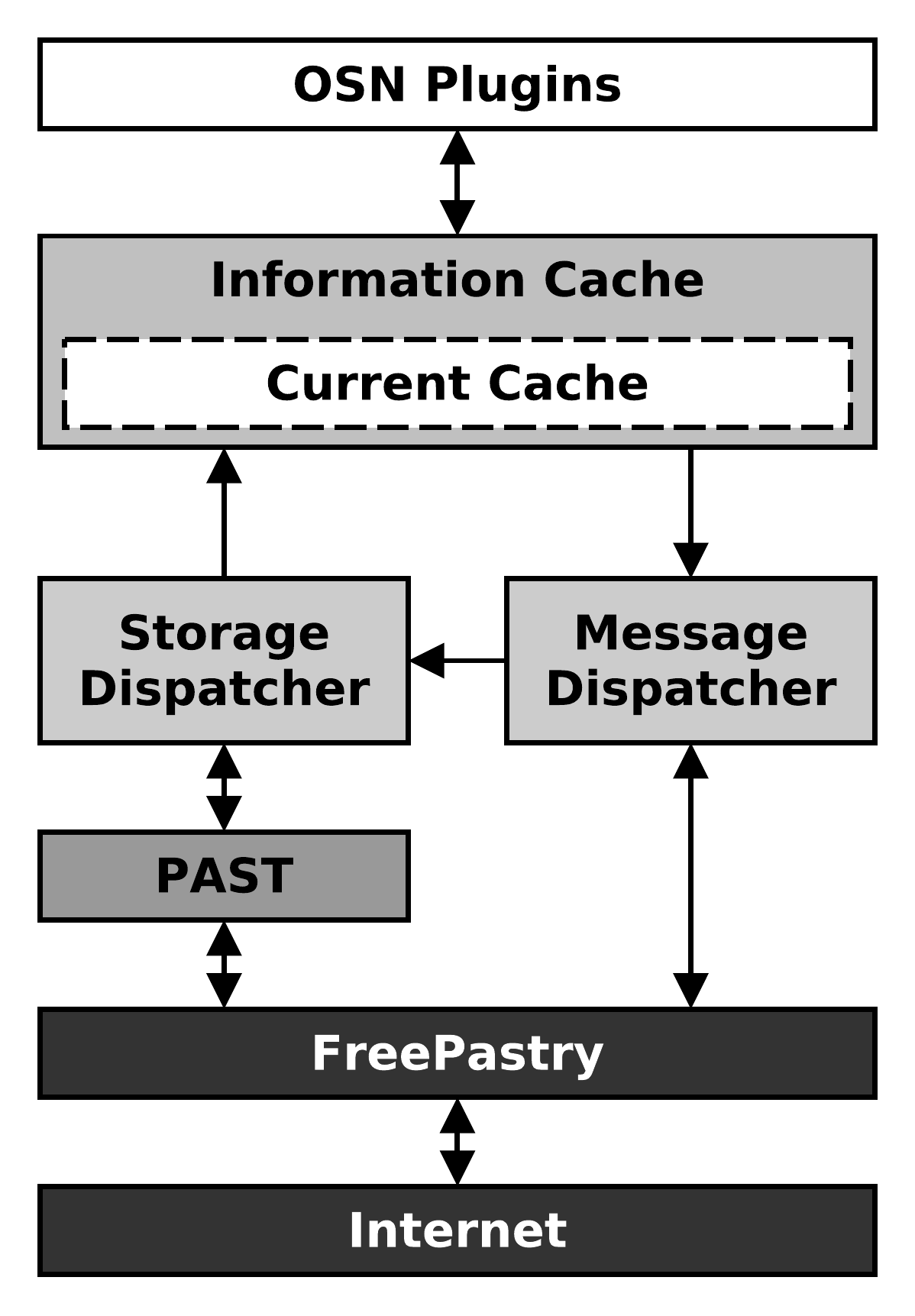}
        \caption{LibreSocial's architecture}
        \label{fig:CurrentDataMgmt}
  \end{minipage}
  \hfill
  \begin{minipage}[b]{0.47\linewidth}
        \centering    
        \includegraphics[width=0.99\linewidth]{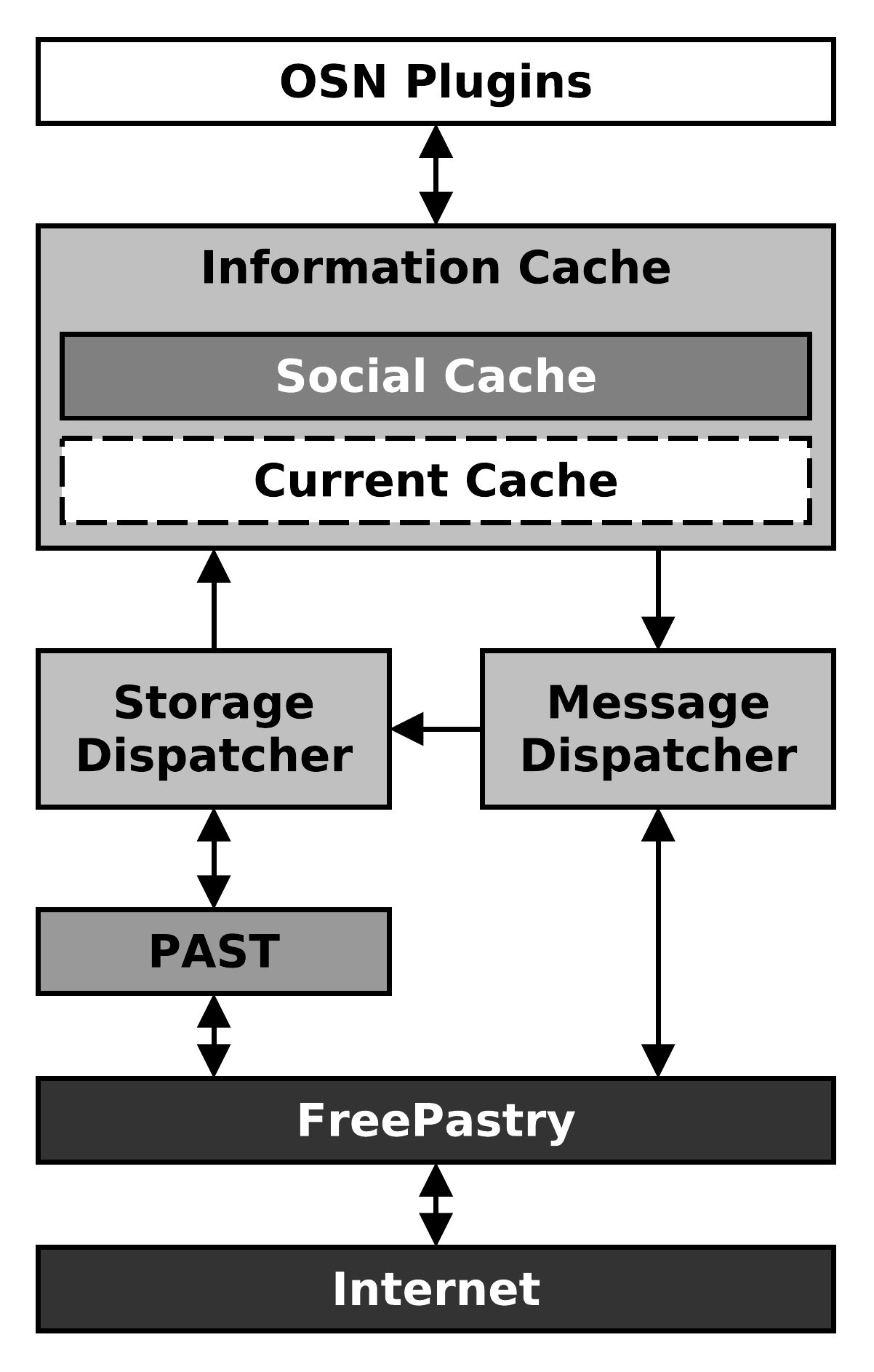}
        \caption{With social cache extension}
        \label{fig:SocialCachePlacement}
  \end{minipage}
\end{figure}

In Fig.~\ref{fig:CurrentDataMgmt}, LibreSocial's data management structure is shown.
In addition to the use of (a heavily secured and extended) PAST for persistence storage, three other components are included for efficient data storage, retrieval and dissemination: \textit{Storage Dispatcher}, \textit{Message Dispatcher}, and \textit{Information Cache}.
The \textit{Storage Dispatcher} provides an interface to interact with the data stored in the overlay network via PAST. It provides additional functionality not previously available in PAST, in particular, the abilities to update data, to remove data and all security elements.
The \textit{Message Dispatcher} is useful in handling system messages and specific events triggered by users such as sending, receiving and replying to friend requests.
Therefore messages sent using the Message Dispatcher must entail a receiving user account along with a payload.
When a message arrives via the Message Dispatcher, the user receives a notification.
In case the user is offline, this message is stored persistently in the network.

The \textit{Information Cache} provides a caching mechanisms for LibreSocial in addition to PAST's internal caching mechanism.
It is designed on top of both the Storage Dispatcher and the Message Dispatcher, building a single layer responsible for data and message access and dissemination for the OSN application.
Messages are accessed via a hash map in accordance to the OSN plugins that the messages are intended for or coming from.
It  hides the delays and complexities due to the asynchronous nature of the system by providing the needed content if available in cache and requesting it otherwise.
The data management process for a node is as follows:
\begin{itemize}
\item
 \textit{First time request from plugins, cache miss}: The Information Cache sends a request to the Storage Dispatcher which performs an overlay lookup.
 Once retrieved, the content is sent to the Information Cache which stores it in the local cache storage, before forwarding it to the requesting plugin.
 Cached content is valid for a fixed amount of time.
\item
 \textit{Consequent requests, cache hit/miss}: A lookup is performed in the local cache.
 If available and valid, it is sent to the requesting plugin.
 If unavailable or not valid, a request is sent to the overlay.
 The received content is updated to the local cache before being delivered it to the requesting plugin.
\item
 \textit{Cache management}: In case the local cache has reached its maximum limit, some of the cache content is deleted based on a ``Least Recently Used'' (LRU) strategy.
\item
 \textit{Adding of new content}: When new content is added by a plugin, the Information Cache first adds it to the local cache before forwarding it to PAST for persistent storage in the network.
\end{itemize}

LibreSocial thus utilizes replication in the DHT for data availability and the request-reply approach for information diffusion.
The request-reply approach is efficient for DHT-based DOSNs but can result in high network traffic due to recurring overlay lookups (for the same content).
Also, in view of the Information Cache freshness, the value set for cache freshness can cause regular overlay lookups if selected too short.
Thus to tackle this, we propose introducing a social cache that employs update dissemination that ensures (relevant) cache data is synchronized after it is updated.
In the next section, we review two DOSNs, Cachet and DiDuSoNet, to gather insight into a possible social caching solution.

\subsection{Cachet vs DiDuSoNet} Cachet~\cite{NJM+12} performs replication in the DHT and uses both reply-request and active dissemination approaches for information diffusion.
Cachet implements social caching using a \textit{pull-push} based gossip algorithm in which a new joining node pulls content from online nodes while a node that generates an update pushes it to other online nodes.
Attribute-based encryption is used to prevent unauthorized data exchanges from the caches when pulling content.
However, in Cachet, the social contact selection algorithm tends to select all online users in a continuous fashion rather than dynamically selecting a few contacts.
This can be achieved easily by collecting information on node interactions by analyzing the social graph structure formed by the social connections but this is not possible in DHT-based DOSNs such as Cachet due to the network topology.

In~\cite{CDG+14}, an epidemic protocol for spreading social updates is suggested, with the links between nodes based on the social interactions.
The protocol uses an egocentric social measure to approximate Betweenness Centrality called the \textit{Weighted Ego Betweenness Centrality (WEBC)} that exploits a weighted graph with the weights referring to the tie strengths between the nodes.
Thus nodes with higher interactions have a higher value of WEBC. This concept is used in DiDuSoNet.
DiDuSoNet~\cite{GAD+16} is a SO-based DOSN that does replication on trusted nodes with the DHT providing indexing services, while relying on request-reply for information diffusion.
Unlike Cachet, DiDuSoNet does not include an advanced caching mechanism but describes a very effective trusted nodes selection mechanisms that makes use of Dunbar-numbers~\cite{Dun98,Dun09} to build an ego network for each user (\textit{ego}) in the network, in which a an ego's network is composed of the trusted nodes.
Dunbar's research on real-life social interactions pointed to the ability of a single person actively maintaining an average of 150 social relationships, referred to as \textit{alters}, at a given time, which is referred to as the \textit{Dunbar number}.
Further,~\cite{AGP13} show that these 150 alters can be divided based on level of closeness around the node into four level such that and each outer layers including the preceding layer, hence 5, 15, 50 and 150.
Using social interaction analysis based on the epidemic protocol in~\cite{CDG+14} to generate the tie strengths, a choice of alters to store the replicas can be made.
This way  the selection of friends to replicate content  done.

\section{A social caching mechanism for LibreSocial}
\label{sec:social_caching_mechanism}


The fundamental structure of LibreSocial's design based on a DHT overlay ensures reliability, scalability and fault-tolerance.
Although LibreSocial assumes no trust among the nodes, not even among ``friends'', it implements key security features that can allow us to develop a social cache structure within the framework, which contains content from nodes that are considered friends hence maintain privacy.
Fig.~\ref{fig:SocialCachePlacement} show the proposed social cache integrated into the Information Cache.
This placement is informed by the fact that the Information Cache interacts directly with the OSN plugins and therefore would provide a platform for the social cache.
We now have to define the social updates, handle lookup requests and selection of suitable users to subscribe to based on social interaction.

\begin{center}
 \begin{figure*}
  \begin{minipage}[b]{0.385\linewidth}
   \centering \includegraphics[width=0.99\linewidth]{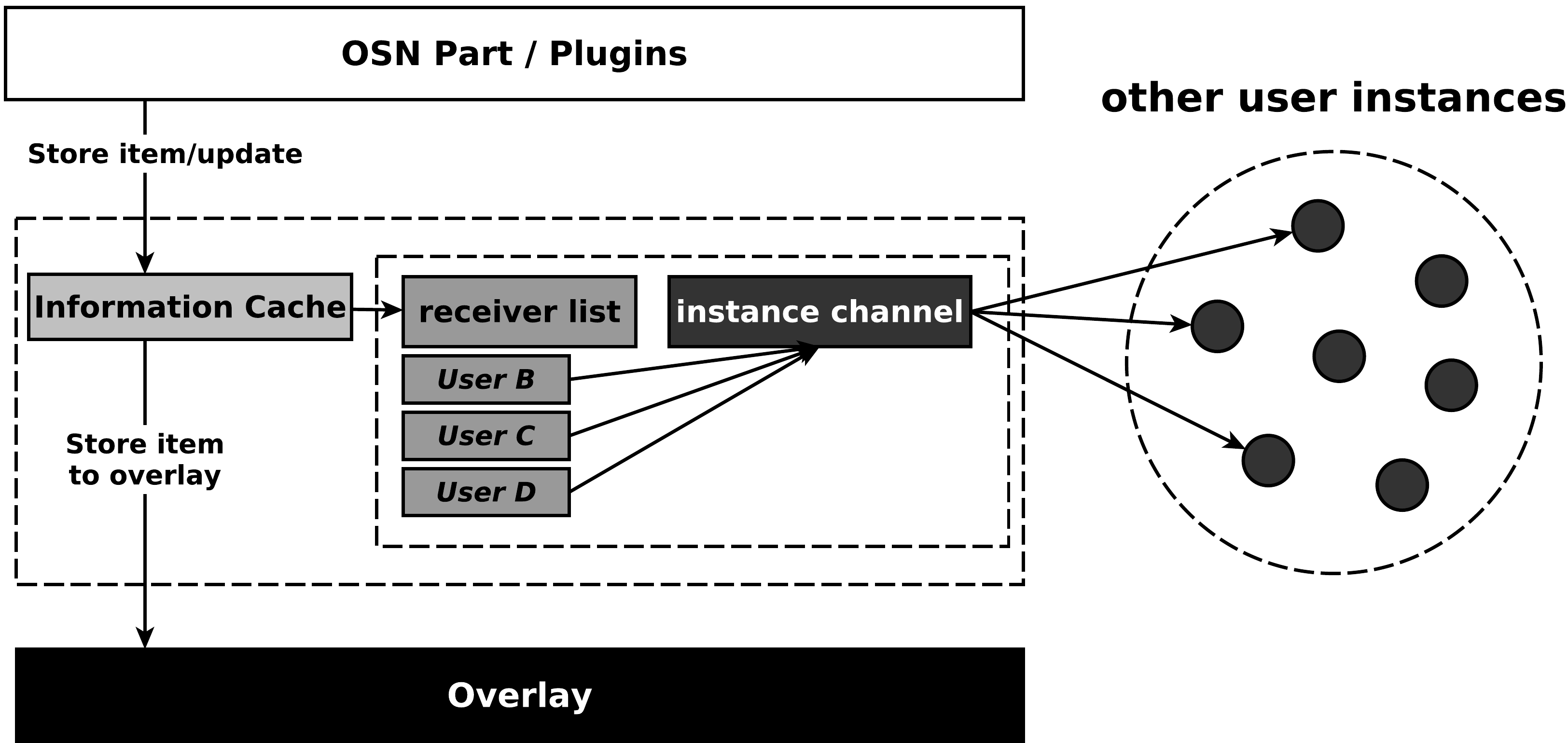}
   \caption{Distribution of social updates}
   \label{fig:social_cache_store}
  \end{minipage}%
  \hfill
  \begin{minipage}[b]{0.385\linewidth}
   \centering \includegraphics[width=0.99\linewidth]{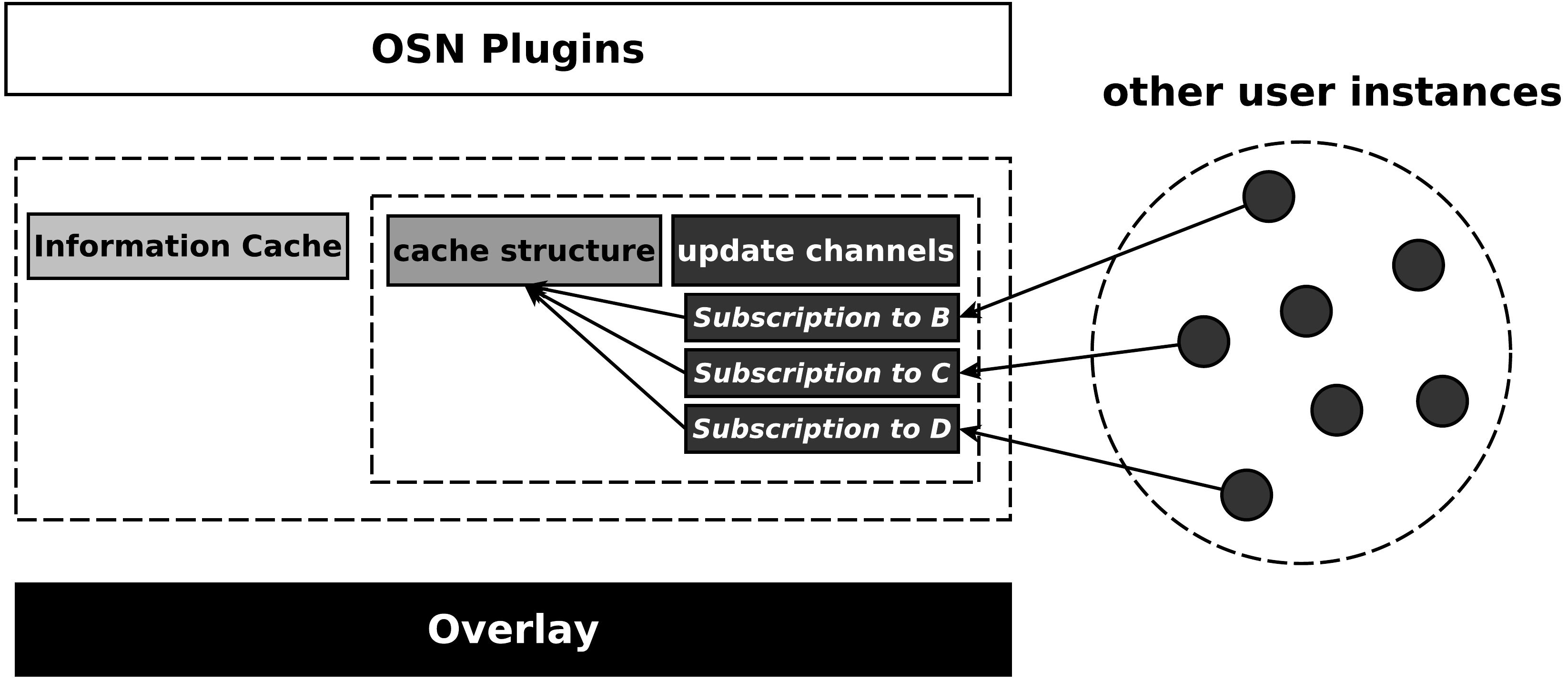}
   \caption{Receiving updates from subscribed channels}
   \label{fig:social_cache_update}
  \end{minipage}
  \hfill
  \begin{minipage}[b]{0.22\linewidth}
   \centering \centering \includegraphics[width=0.99\linewidth]{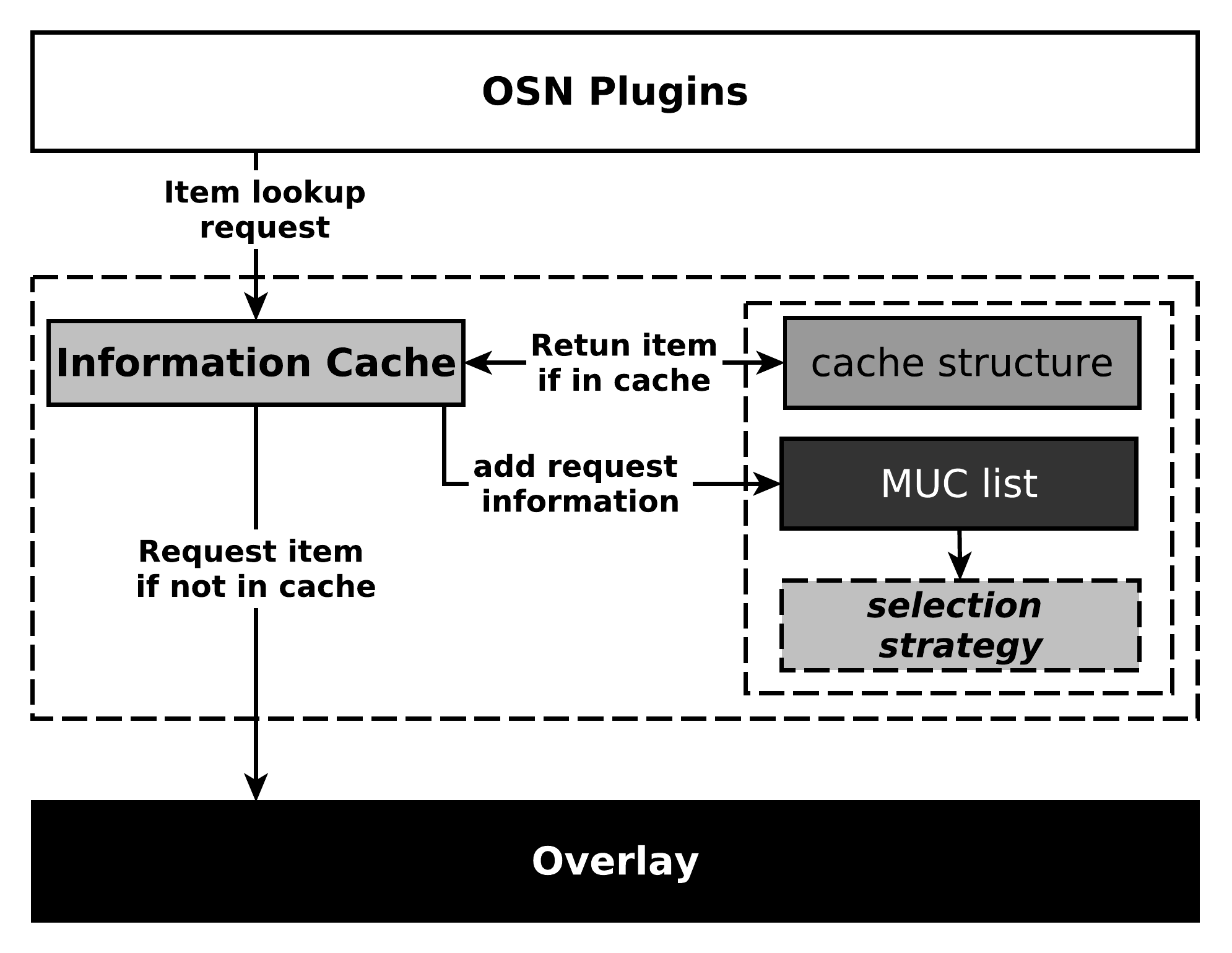}
   \caption{Lookup request using the social cache}
   \label{fig:social_cache_lookup}
  \end{minipage}
 \end{figure*}
\end{center}

\subsection{Social updates} In LibreSocial the cached data is not actively re-synchronized with the original data, hence cached data consistency is not guaranteed but the maximum age of the data is limited.
By use of \textit{social updates} we can actively disseminate changed content to selected users for the case. 
Once they would request the content (as predicted) it will be already available locally. 
Performing social updates is characterized by two aspects: \textit{detection and distribution of content changes}, and \textit{receiving and storing the updates}.
We consider each aspect carefully

\subsubsection{Detection \&
 distribution of content changes} The social update includes content modified by its originator disseminated via a messaging channel, called an \textit{instance channel}, to a defined group of users listening to that channel.
Each social caching instance maintains a list of selected users subscribed to the upcoming content modifications of this instance called \textit{receiver list}.
When a user performs any content modifications that updates the data in the DHT, the social cache also transmits a social update for each of the subscribed users stored in its receiver list.
This process is illustrated in Fig.~\ref{fig:social_cache_store}.
This guarantees that users subscribed to upcoming social updates of a particular user's instance channel receive content updates.

\begin{algorithm}[t]
 \tiny 
 \DontPrintSemicolon 
 \SetAlgoLined 
 \DontPrintSemicolon 
 \SetAlgoLined 
 \SetKwData{Username}{username} 
 \SetKwData{UpdateLimit}{updateLimit} 
 \SetKwData{MUCList}{MUC\_List} 
 \SetKwData{SelectionStrategy}{selectionStrategy} 
 \SetKwArray{SubscriptionChannels}{subscriptionChannels} 
 \SetKwFunction{GetUsername}{getUsername} 
 \SetKwFunction{SizeOfList}{sizeOfList} 
 \SetKwFunction{Subscribe}{subscribe2Updates} 
 \SetKwFunction{RandomSelectionStrategy}{RandomSelectionStrategy} 
 \SetKwFunction{TrendSelectionStrategy}{TrendSelectionStrategy} 
 \SetKwFunction{SocialScoreSelectionStrategy}{SocialScoreSelectionStrategy}
 \SetKwIF{If}{ElseIf}{Else}{if}{then}{else if}{else}{endif} 
 \SetKwInOut{Input}{Input} 
 \SetKwInOut{Output}{Output} 
 \Input{\textit{storageKey},  \textit{MUC\_ List}  \textit{subscriptionChannels}, \textit{updateLimit}, \textit{selectionStrategy}} 
 \Output{Subscripion list based on selection strategy} 
 \Username $\gets$ \GetUsername{$storageKey$};\ 
    \If{\SelectionStrategy ==   $random$}{ 
        \If{\Username   $\in   $ \MUCList}{ 
            \RandomSelectionStrategy{\Username}\; 
        }
    } 
    \ElseIf{\SelectionStrategy ==   $trend$}{ 
        \If{\Username    $\in   $ \MUCList}{ 
            \textit{numOfChannels}   $\gets$ \SizeOfList{\SubscriptionChannels}\;
            \If{$numOfChannels <    $ \UpdateLimit}{ 
                \Subscribe{\Username};\ 
            } 
            \Else{ \TrendSelectionStrategy{\Username} } 
        } 
    } 
    \ElseIf{\SelectionStrategy ==  $socialScore$}{ 
        \If{\Username   $\in   $ \MUCList}{ 
            \textit{numOfChannels}   $\gets$ \SizeOfList{\SubscriptionChannels}\;
            \If{$numOfChannels <     $ \UpdateLimit}{ 
                \Subscribe{\Username};\ 
            } 
            \Else{ \SocialScoreSelectionStrategy{\Username} } 
        } 
    }
 \caption{Information Cache Lookup}
 \label{alg:InformationCacheLookup}
\end{algorithm}

\subsubsection{Receiving \&
 storing update} Each social caching instance handles its own subscriptions to other users' instance channels.
This is shown in Fig.~\ref{fig:social_cache_update}.
The social cache listens to the instance channels for updates and stores them into a data structure called \textit{update channels}.
Every time a social update is populated through one of this update channels, the social cache receives the stored information from the update message and stores it in its own caching data structure.
The caching structure in the social cache is organized in two-layers and is shown in figure \ref{fig:social_cache_cache_structure}.
For each subscribed user, received information is stored based on its \textit{StorageKey} and every social update contains information regarding the subscribed user and the particular StorageKey.
Previous information is overwritten by a newly incoming social update.
By this, availability and consistency for all stored items in the social cache is assured.


The next issue of concern are now the extensions to the Information Cache to support the lookup process using the received social updates instead of performing overlay requests.
This is considered next.

\subsection{Handling lookup requests} The social cache is integrated to work with the lookup process within the Information Cache.
Lookup requests contain the username of the information requestor as well as the StorageKey of the needed information.
Upon receipt of a request from an OSN plugin, the Information Caches first tries to locate the content in the caching structure of the social cache by matching cached data against the StorageKey.
If available, the content is sent to the requesting plugin.
If the requested content is not found in the social cache, then a request is issued to the DHT.
This process is shown in Fig.~\ref{fig:social_cache_lookup}.

\begin{center}
 \begin{figure}
\vspace{-0.35cm}
 \begin{minipage}[b]{0.59\linewidth}
   \centering \centering \includegraphics[width=0.99\linewidth]{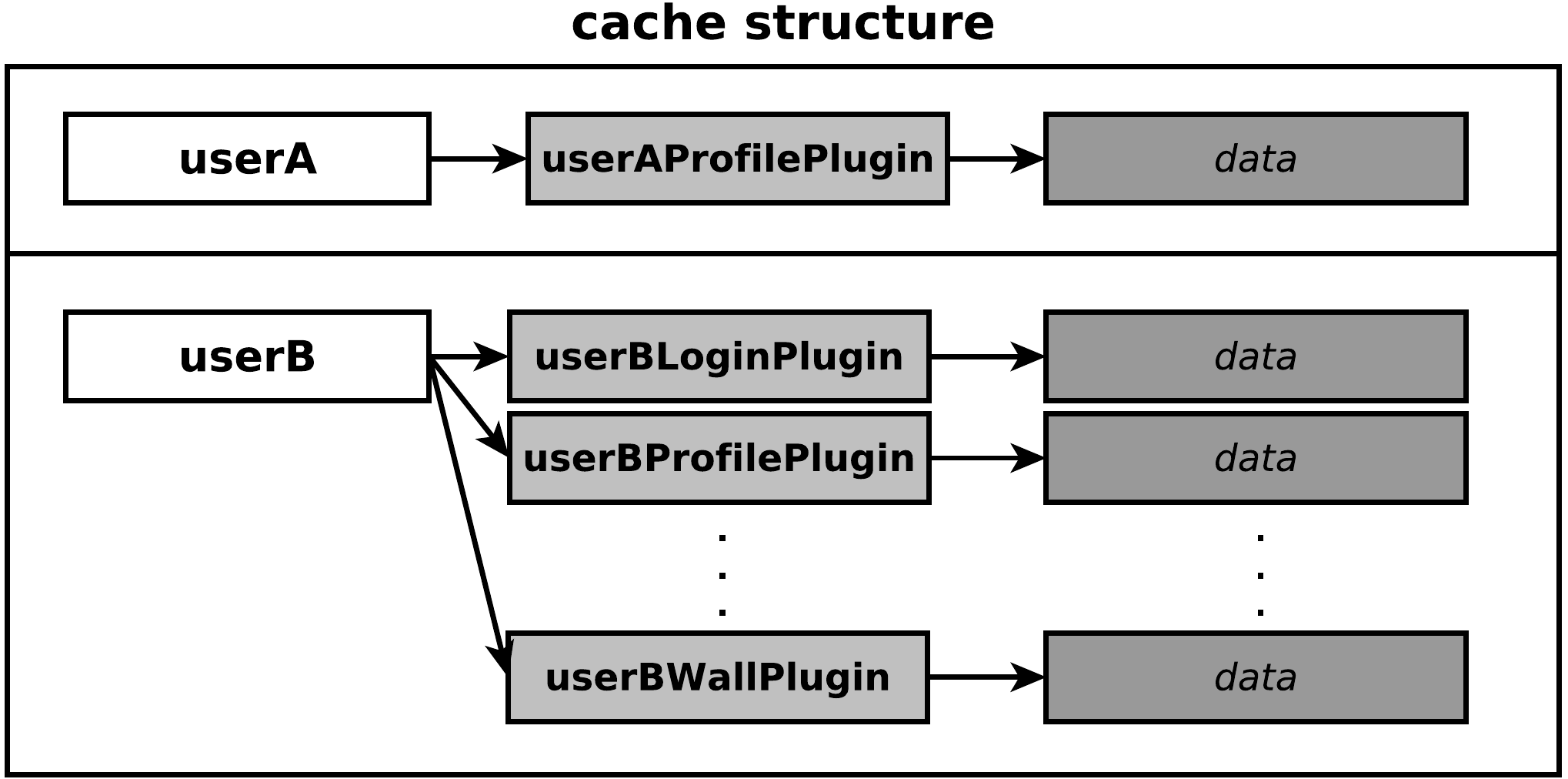}
   \caption{Social cache with two nested   \mbox{abstraction} levels}
   \label{fig:social_cache_cache_structure}
  \end{minipage}%
  \hfill
  \begin{minipage}[b]{0.40\linewidth}
   \centering \centering \includegraphics[width=0.99\linewidth]{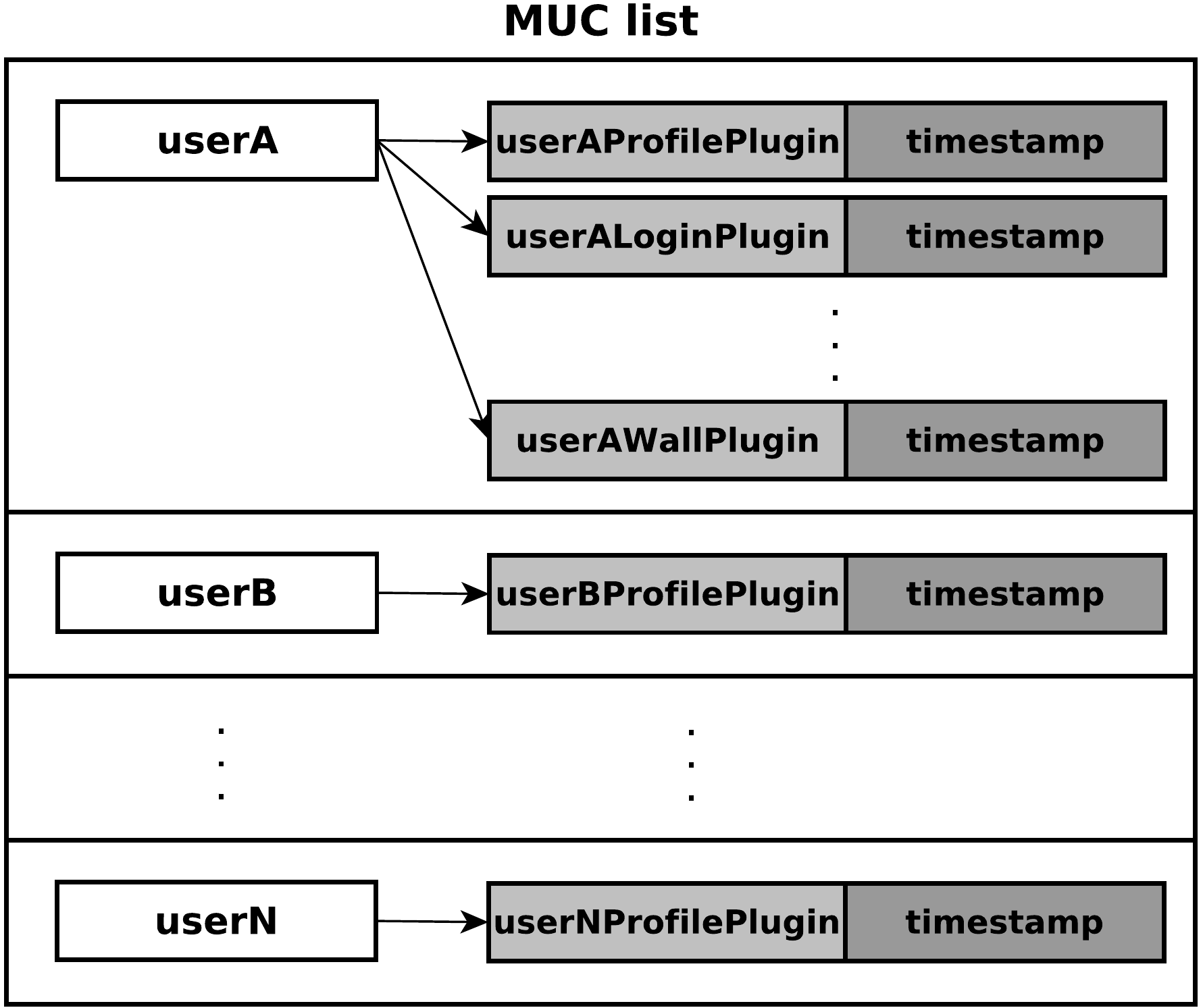}
   \caption{MUC list structure}
   \label{fig:muc_list_social_score}
  \end{minipage}
 \end{figure}
 \vspace{-0.3cm}
 \end{center}
  \begin{algorithm}
        \tiny 
        \DontPrintSemicolon 
        \SetAlgoLined 
        \SetKwData{UserName}{userName} 
        \SetKwArray{SubscriptionChannels}{subscriptionChannels} 
        \SetKwData{UpdateLimit}{updateLimit} 
        \SetKwFunction{SizeOfList}{sizeOfList} 
        \SetKwFunction{GetRandomNum}{getRandomNum} 
        \SetKwFunction{Subscribe}{subscribe2Updates} 
        \SetKwFunction{Unsubscribe}{unsubscribe2Updates} 
        \SetKwInOut{Input}{Input} 
        \SetKwInOut{Output}{Output} 
        \Input{\textit{userName}, \textit{subscriptionChannels}, \textit{updateLimit}} 
        
        \Output{Randomly selected replication list}
        $numOfChannels \gets$ \SizeOfList{\SubscriptionChannels}\;
        \uIf{$numOfChannels  <$ \UpdateLimit}{ \Subscribe(\UserName)\; } 
        \uElse {
        $randomIndex   \gets$ \GetRandomNum{0, $numOfChannels$-1}\;
        \Unsubscribe{\SubscriptionChannels{$randomIndex$}}\; 
        \Subscribe{\UserName}; }
        \caption{Random Selection Strategy}
        \label{alg:RandomStrategy}
    \end{algorithm}

\subsection{Selecting suitable users based on social interaction} 
The next major step is the selection strategy for subscriptions.
The \textit{selection strategy} is the method used to filter and select a set of suitable users to which an instance then subscribes to so as to receive social updates.
The suitable subscriptions should ideally be those that will be accessed frequently in the future.
Because the only way to achieve a high ratio of cache hits for the social caching mechanism is to ensure that data accessed in future mainly belongs to users that are currently subscribed to. 
The selection of those users is a crucial component of the social caching mechanism.

To make an informed decision on subscriptions to make, the social cache monitors the interactions an instance makes with other users.
We further make the restriction that the social cache does not monitor the application plugins but only keeps track of all incoming requests received by the Information Cache from the OSN plugins without consideration of the source of the final response to the request.
For every request, details on the user making the request, the type of request and the timestamp of the request are stored.
This information is stored in a separate data structure called \textit{most-used-contacts} (MUC) list, from which suitable subscriptions are to be made.
In Fig.~\ref{fig:muc_list_social_score} a representation of such a MUC list is shown.
This data is independent from the actual social behavior of the users as viewed from the social network, but it nevertheless offers information about how frequently and in which way the own instance is interacting with other users.

The MUC list is the basis for the selection strategy that is used to choose the most suitable users to subscribe to.
The selection process is triggered in fixed intervals called \textit{update intervals}.
Let $m$ be a predefined maximum number of tracked lookup requests and $n$ the number of parallel maintainable update channels with $n<m$. 
After $m$ lookups, $n$ candidates are chosen who are deemed most suitable to subscribe to during the next update interval. 
We propose three selection strategies for these $n$ candidates, namely \textit{random}, \textit{trend}, and \textit{social score} selection strategy which are further described.
The general Information Cache lookup is shown in Algorithm~\ref{alg:InformationCacheLookup}.

\subsubsection{Random selection strategy} The random selection strategy is shown in Algorithm~\ref{alg:RandomStrategy}.
In this strategy, the user subscribes to every new user who is tracked in lookup process until the limit of
$n$ candidates to subscribe to is reached.
If that limit is reached and a new user is discovered through the MUC list process, a currently subscribed user is randomly chosen to unsubscribe from and the new subscription is added.
In this selection strategy the MUC list is only used to monitor known and unknown user.
The stored lookup information does not affect the selection's decision in any way.
An unsubscription of a user also leads to the removal of this user from the MUC list.
Therefore the number of stored users in the MUC List is the same as the currently maintained parallel subscriptions.

\subsubsection{Trend selection strategy} The trend selection strategy is given in Algorithm~\ref{alg:TrendStrategy}.
It utilizes the MUC list in a different way from the random selection strategy.
Once the required
$m$ lookups are achieved, the tracked users stored in the MUC list are ranked based on the number of performed lookups from highest to lowest.
Then the first
$n$ users are chosen, which are assumed to have the most interaction with the user over the current lookup interval.
During the next lookup interval, subscriptions are then sent to these
$n$ users as long as there are no existing subscriptions.
In case of existing subscriptions, subscribed users not in the generated ranking are unsubscribed from.
Thereafter, the MUC List is cleared for the new lookup interval.

\begin{algorithm}[t]
 \tiny 
 \DontPrintSemicolon 
 \SetAlgoLined 
 \DontPrintSemicolon 
 \SetAlgoLined 
 \SetKwData{UpdateLimit}{updateLimit} 
 \SetKwData{MUCList}{MUC\_List} 
 \SetKwArray{SubscriptionChannels}{subscriptionChannels} 
 \SetKwArray{TrendArray}{trendArray} 
 \SetKwArray{NewSubscriptions}{newSubscriptions} 
 \SetKwArray{OldSubscriptions}{oldSubscriptions} 
 \SetKwFunction{SortByDescending}{sortByDescending} 
 \SetKwFunction{Subscribe}{subscribe2Updates} 
 \SetKwFunction{Unsubscribe}{unsubscribe2Updates} 
 \SetKwInOut{Input}{Input} 
 \SetKwInOut{Output}{Output} 
 \Input{\textit{username}, \textit{subscriptionChannels}, \textit{updateLimit}} 
 \Output{Replication list based on trends} 
 \TrendArray  $\gets \varnothing$\;
 \ForEach{$user \in  $ \MUCList}{ 
        \TrendArray  $\gets$ number of requests by  $user$\; 
 } 
 \SortByDescending{\TrendArray}\; 
 \NewSubscriptions $\gets \varnothing$\;
  $i \gets 0$\;
 \While{$i <  updateLimit$}{ 
        \NewSubscriptions  $\gets$ \TrendArray{$i$}\;
        $i=i+1$\; 
 } 
 \OldSubscriptions  $\gets \varnothing$\; 
 \OldSubscriptions $\gets subscriptionChannels$ \;
 \ForEach{$user \in  ($\NewSubscriptions  $\cap  \neg$ \OldSubscriptions$)$}{
        \Subscribe{$user$}\; 
 }
 \ForEach{$user \in  (\neg$ \NewSubscriptions  $\cap$ \OldSubscriptions$)$}{
        \Unsubscribe{$user$}\; 
 } 
 \MUCList $\gets \varnothing$
 \caption{Trend Selection Strategy}
 \label{alg:TrendStrategy}
\end{algorithm}

\subsubsection{Social score selection strategy} The strategy is shown in Algorithm~\ref{alg:SocialScoreStrategy}.
It follows the idea of using a \textit{SocialScore} based on Dunbar's~\cite{Dun98} social interaction model,  as we sketched for DiDuSoNet~\cite{GAD+16}.
While DiDuSoNet is based on a social overlay and uses a mixture of overlay connection and interaction information to determine trusted nodes for replication based on social closeness, we adopt the mechanism in the scope of the DHT-based nature of LibreSocial to express strong social connections between a user with other users.
This strategy deviates from the trend selection strategy based on how the ranking of users in the MUC list during the subscription is done, in addition to not clearing the MUC list during a new lookup interval.
When the lookup interval of
$m$ lookups is reached, for each user in the MUC list a social score is calculated.
The social score for a user
$x$ is calculated as shown in Equation~\ref{eq:social_score_ls}.
It is based on the value of the \textit{TieStrength} and the \textit{MediumInteractionLength} (MIL) where
$\alpha$ and
$\beta$ are weights that are defined by the users based on relevance given to each criteria.
\begin{equation}
 \scriptsize
 \begin{aligned}
  \mathrm{SocialScore_{LS}(x)} = \alpha * \mathrm{TieStrength(x)} + \beta * \mathrm{MIL(x)}
  \label{eq:social_score_ls}
 \end{aligned}
\end{equation}


The \textit{TieStrength} as defined by~\cite{GAD+16} is a numerical value that helps to define the relationship between two users and can be calculating by considering factors such as frequency of contact between users, number of likes, posts, tags, comments and so on~\cite{AGP13}.
It is obtained by summing all tracked interactions for a user
$x$ divided by the total amount of all monitored interactions
$I$ so as to normalize it based on the total number of interactions by user
$x$.
\begin{equation}
 \scriptsize
 \begin{aligned}
  \mathrm{TieStrength(x)} = \displaystyle\dfrac{\displaystyle\sum_{i=1}^{n \in I(x)}\mathrm{weight(interaction_i)}}{|I|}
  \label{eq:tie_strength_ls}
 \end{aligned}
\end{equation}
The metric MIL, shown in Equation~\ref{eq:mil_ls}, is the average duration of interactions between two users and is introduced in place of the ranking of users done for the trend selection strategy.
It is calculated by taking the medium time span between all tracked interactions for a user
$x$ and in relation to the time period between the first recorded interaction and the current timestamp available for user
$x$ at the time of the calculation.
\begin{equation}
 \scriptsize
 \begin{aligned}
  \mathrm{MIL(x)} = \displaystyle\dfrac{\displaystyle\sum_{i=1}^{n}\left[\tfrac{(T_i(x) - T_{i-1}(x))}{n-1}\right]}{T_{\mathrm{now}} - T_\mathrm{0(x)}}
  \label{eq:mil_ls}
 \end{aligned}
\end{equation}
By combining the values of \textit{TieStrength} and \textit{MIL} it is now possible to obtain a relation between the number of interactions against length of interactions for every monitored user.
Hence, the
$SocialScore_{LS}$ ranks a user with more frequent interactions over a longer period of time higher, than a user with fewer interactions and shorter period. 
This mechanism ensures consistent cache entries, takes into account the social component of an OSN, dynamically adopts to changes of users preferences, and is fully working without undermining the existing processes of the given application.
%

Nevertheless, there are some special cases in which the social caching mechanism has conceptual weaknesses or limits.
Those weaknesses and limits need to be considered and can be solved by additional processes which are discussed in the following.

\begin{algorithm}[t]
 \tiny 
 \DontPrintSemicolon 
 \SetAlgoLined 
 \DontPrintSemicolon 
 \SetAlgoLined 
 \SetKwData{UpdateLimit}{updateLimit} 
 \SetKwData{MUCList}{MUC\_List} 
 \SetKwArray{SubscriptionChannels}{subscriptionChannels} 
 \SetKwArray{SocialScoreArray}{socialScoreArray} 
 \SetKwArray{NewSubscriptions}{newSubscriptions} 
 \SetKwArray{OldSubscriptions}{oldSubscriptions} 
 \SetKwFunction{CalculateSocialScore}{calculateSocialScore} 
 \SetKwFunction{SortByDescending}{sortByDescending} 
 \SetKwFunction{Subscribe}{subscribe2Updates} 
 \SetKwFunction{Unsubscribe}{unsubscribe2Updates} 
 \SetKwFor{ForEach}{foreach}{do}{endfch} 
 \SetKwInOut{Input}{Input} 
 \SetKwInOut{Output}{Output} 
 \Input{ \textit{username}, \textit{subscriptionChannels}, \textit{updateLimit}} 
 \Output{Replication list based on trends} 
 \SocialScoreArray  $\gets \varnothing$\;
 \ForEach{$user \in $ \MUCList}{
    $socialScore   \gets$ \CalculateSocialScore{$user$}\; 
    \SocialScoreArray  $\gets  socialScore$\; 
 } 
 \SortByDescending{\SocialScoreArray}\; 
 \NewSubscriptions $\gets \varnothing$\;
 $i \gets 0$\;
 \While{$i <  updateLimit$}{
    $newSubscriptions  \gets$ \SocialScoreArray{$i$}\;
    $i=i+1$\; 
 } 
 \OldSubscriptions $\gets \varnothing$\; 
 \OldSubscriptions $\gets subscriptionChannels$ \;
 \ForEach{$user \in  ($\NewSubscriptions  $\cap  \neg$ \OldSubscriptions$)$}{
    \Subscribe{$user$}\; 
 }
 \ForEach{$user \in  (\neg$ \NewSubscriptions  $\cap$ \OldSubscriptions$)$}{
    \Unsubscribe{$user$}\; 
 }
 \caption{Social Score Selection Strategy}
 \label{alg:SocialScoreStrategy}
\end{algorithm}

\subsection{Addressing constraints in social caching}
With the combination of the proposed concepts of social updates, the tracking of interactions by the MUC list, and the selection of subscriptions by selection strategies, we have created a caching mechanism, which uses active dissemination of newly updated information to a fixed set of users. 
This proactive caching is optional and allows for a quicker, as local, resolving of requests at the LibreSocial instances of friends. 
However, there are some  cases in which the social caching mechanism presents conceptual weaknesses or limits which need to be addressed via additional processes which we highlight.

\subsubsection{Recurrent overlay requests for own content} 
Sometimes, lookups may not result in cache hits thus leading to overlay requests, negating the use of the social cache.
This can occur when a user does not have his/her friends' content in the social cache, but may need to also access this content often and may therefore result in recurrent overlay requests.
As changes to the content are initiated by the OSN plugins which are on top of the social caching process, we can easily store them in an additional data structure and serve the requests from the social cache instead.
With this data structure we do not undermine the consistency of the stored information but reduce the needed overlay requests by a significant amount.

\subsubsection{Social bootstrapping} When one of the current subscription hardly make changes to content, there will be no social updates received from them.
Since only social updates lead to cache entries, requests for this user will always cause recurrent overlay request.
To tackle this case we introduced the concept of \textit{social bootstrapping}.
Every time a subscription between users is performed, the user receiving the subscription request sends a message to the subscriber to inform him/her to receive social updates from that point onward.
The user being subscribed to then answers with all of their own cached content which is inserted into the caching structure of the subscribing user.
The consistency of cached content is not affected by this process, because a change in the initially sent content will lead to a social update and which will overwrite the initial content.

\subsubsection{Limitation of active subscription and Dunbar-number} So far it has been assumed that there is a limitation placed on the number of parallel active subscriptions without giving an actual limiting value.
As the social cache is designed as additional mechanism in a given DHT-based overlay, the number of connections between users needs to be restricted to prevent establishment of a second overlay.
This limiting value is based on the \textit{Dunbar number}.
Because the maximum amount of maintainable contacts is at about 150 according to Dunbar~\cite{Dun98}, we also restrict the MUC list to store information for a maximum of 150 distinct users.
When this limit is reached, the users with the least rank according to the selection strategy is removed to create space for newly monitored users.

\section{Experimental Setup}
\label{sec:evaluation}
Evaluation of the social caching mechanism is the process of benchmarking the overall performance of LibreSocial in the face of the new mechanism.
A benchmark is in general a tuple of \textit{quality attributes} (or \textit{properties})
$Q$, \textit{metrics}
$M$, and test scenarios
$S$~\cite{KGK+08}.
In the discussion that follows, we describe each of these in view of the social caching mechanism evaluation.
Thereafter, the results are presented and discussed.

\subsection{Quality properties and metrics} While the metrics focus on only one attribute of the system or mechanisms within the test scenario, the quality attributes generally describe the same system or mechanism characteristics while taking into account several metric measurements.
Thus, with the use of a set of quality metrics, useful statements about the general characteristics can be made which describe the overall quality of the system.
For purposes of analyzing the social caching mechanism, the quality properties chosen were performance and efficiency.
This are further considered.

\subsubsection{Performance} 
This characterizes the system in terms of how it responds, its throughput and the validity of the results as a consequence of a particular workload within the bounds of the test environment.
In the case of the proposed social caching mechanism, the overall objective is a reduction in overlay requests and an increase in cache-based responses to the content requested.
Thus by collecting the overall requests against the answered requests from the cache and from the overlay, we can calculate the metric \textit{cache hit ratio} as shown in Equation~\ref{eq:CacheHitRatio}, which helps in quantifying the performance.
The higher the value of the cache hit ratio, the better the performance
\begin{equation}
 \scriptsize
 \begin{aligned}
  Cache\_hit\_ratio = \displaystyle\dfrac{Cache\_replies}{Total\_replies}
  \label{eq:CacheHitRatio}
 \end{aligned}
\end{equation}

\subsubsection{Efficiency} 
This is defined as the ratio between performance and costs when considering a particular task that the system is undertaking.
For the proposed social caching mechanism, we distinguish between two types, that is \textit{local efficiency} and \textit{overall efficiency}.
\begin{itemize}
\item
 \textbf{Local efficiency}: With this, we desire to find out how certain resources in the local environment, in this case, the local node, is utilized.
 As more content is to be stored locally in the cache, it is expected that more system resource will be utilized.
 We desire to ensure that the local resources are not used up, such as memory as well as processing power.
 Therefore, we monitor the number of objects cached  in order to find out the required memory to achieve a desired cache hit ratio, as well as the average MUC list size, and the number of subscription and unsubscription processes.
 Generally, smaller values observed in all these metrics in combination are an indicator of good local efficiency.
\item
 \textbf{Overlay efficiency}: Since we are working in a distributed environment, we would like to find out the overall workload generated by the system and its effects on several aspects of the system, and in particular, the overlay.
 Thus we keep track of the number of messages in the overlay, the number of PAST and Pastry data operations, and average storage used to maintain overlay processes.
 Therefore, we observe different aspects of the LibreSocial's overlay like the number of messages passed through it, the number of Pastry/PAST data operations, and the amount of storage used to maintain the overlay processes in each node.
\end{itemize}

Having established the quality properties and their relevant metrics, we now consider the test scenario.

\subsection{Test scenario}
\label{subsec:test_scenario}
The test scenario constitutes the environment in which a test is being undertaken, involving all the system parameters, the workload parameters and the workload itself.
Different test scenarios affect the quality of the system in different ways.
To effectively evaluate the proposed social caching mechanism, we take into consideration the number of interactions and the time interval occurring between these interactions.
This points to the need of a more realistic test scenario.
For this, three suitable OSN data sets were considered, which provide realistic information about the number of interactions as well as time intervals between interactions.
The data sets are the Facebook'09 \cite{VMC+09}, SNAP'14 \cite{CDG+14a}, and Facebook'14 \cite{GAD+16}.
These data sets are also ego-based, contain information about the relations and are all based in analysis of Facebook.
The information that can be gleaned from these data sets is shown in Table~\ref{table:fb_data_sets}.

The SNAP'14 data set only include information about the friend relations and the user profiles, and the data has no information about interactions between users nor interaction times through timestamps and hence is unsuitable.
The Facebook'14 data set provides information about friend relations and different types of interactions (such as wall posts, likes, tagged photos) as well as relevant timestamp indicating time of interactions.
Unfortunately due to privacy concerns, this data set could not be available at the time of testing.
Therefore, Facebook'09 data was chosen, as it was available and fulfills the need for a data set that reveals information about friend connections between users, interactions between the users as well as the timestamps for the interactions.
The Facebook'09 data set is further analyzed so as to obtain relevant real world statistical information, shown in Table~\ref{table:data_set_analysis}, that would be useful in designing the test scenario.

\begin{table*}
 \begin{minipage}[b]{0.4\textwidth}
  \centering
  \tiny
  \caption{Available data sets with relevant features}
  \label{table:fb_data_sets}
  \begin{tabular}{|l|l|l|l|l|l|}
   \hline \rule{0pt}{2ex}\multirow{2}{*}{\textit{Data set}\textbf{}} &
   \multirow{2}{*}{\textbf{Year}} &
   \multicolumn{4}{c|}{\textbf{Available data}} \\
   \cline{3-6} \rule{0pt}{2ex} &
   &
   Timestamps &
   Friends &
   Interactions &
   Profile \\
   \hline \rule{0pt}{2ex}\textit{Facebook'09} &
   2009 &
   \cmark &
   \cmark &
   \cmark &
   \\
   \hline \rule{0pt}{2ex}\textit{SNAP'14} &
   2014 &
   &
   \cmark &
   &
   \cmark \\
   \hline \rule{0pt}{2ex}\textit{Facebook'14} &
   2014 &
   \cmark &
   \cmark &
   \cmark &
   ?
   \\
   \hline
  \end{tabular}
 \end{minipage}%
 \hfill
 \begin{minipage}[b]{0.30\textwidth}
  \centering
  \tiny
  \caption{Available features in Facebook'09 data set}
  \label{table:data_set_analysis}
  \begin{tabular}{|l|l|}
   \hline \rule{0pt}{2ex}Total \# of egos &
   60102 \\
   \hline \rule{0pt}{2ex}Total \# of alters &
   1545690 \\
   \hline \rule{0pt}{2ex}Average \# of alters &
   25.7177 \\
   \hline \rule{0pt}{2ex}Average TS/new friend request &
   36.7332 days \\
   \hline \rule{0pt}{2ex}Average TS/new interaction &
   43.0402 days \\
   \hline \rule{0pt}{2ex}Experiment time span &
   869.458 days\\
   \hline 
  \end{tabular}
 \end{minipage}%
 \hfill
  \begin{minipage}[b]{0.30\textwidth}
  \centering
  \tiny
  \caption{Experimental parameters and conditions}
  \label{table:test_setup_parameters}
  \begin{tabular}{|l|l|c|}
   \hline \rule{0pt}{2ex}\textbf{Test Environment} &
   LibreSocial instances &
   64 \\
   \rule{0pt}{2ex}&
   Physical machines &
   8 \\
   \hline \rule{0pt}{2ex}\textbf{Application settings} &
   Social update channels &
   15 \\
   \rule{0pt}{2ex}&
   MUC list size &
   150 \\
   \rule{0pt}{2ex}&
   Update interval &
   50 sec.\\
   \rule{0pt}{2ex}&
   Social bootstrapping &
   enabled \\
   \hline \rule{0pt}{2ex}\textbf{Duration of Experiment} &
   Selection strategies &
   6 hours \\
   \rule{0pt}{2ex}&
   Caching approaches &
   2 days \\
   \hline
  \end{tabular}
 \end{minipage}
\end{table*}

\begin{table*}
 \begin{minipage}[b]{0.30\textwidth}
  \centering
  \tiny
  \caption{Answered requests comparison based on selection strategies}
  \label{table:eval_ss_perf}
  \begin{tabular}{|p{0.5cm}|p{0.5cm}|p{0.5cm}|p{0.5cm}|p{0.6cm}|}
   \hline \rule{0pt}{2ex} \multirow{2}{*}{\begin{minipage}{1.0cm}\textbf{Strategy}\end{minipage}} &
   \multicolumn{3}{c|}{\begin{minipage}{1.5cm}\textbf{Requests answered}\end{minipage}} &
   \multirow{2}{*}{\textbf{\begin{minipage}{0.7cm}Cache hit ratio\end{minipage}}} \\
   \cline{2-4} \rule{0pt}{2ex} &
   \textit{Cache} &
   \textit{Overlay} &
   \textit{Total} &
   \\
   \hline \rule{0pt}{2ex}Random &
   635663 &
   33032 &
   669476 &
   0.949 \\
   \hline \rule{0pt}{2ex}Trend &
   387253 &
   26987 &
   414971 &
   0.933 \\
   \hline \rule{0pt}{2ex}Social Score &
   1054618 &
   85355 &
   1140900 &
   0.924 \\
   \hline
  \end{tabular}
 \end{minipage}%
\hfill
 \begin{minipage}[b]{0.33\textwidth}
  \centering
  \tiny
  \caption{Distribution of request responses}
  \label{table:eval_cs_perf}
  \begin{tabular}{|p{0.6cm}|p{0.5cm}|p{0.5cm}|p{0.5cm}|p{0.5cm}|p{0.6cm}|}
   \hline \rule{0pt}{2ex} \multirow{3}{*}{\begin{minipage}{0.7cm}\textbf{Cache setup}\end{minipage}} &
   \multicolumn{4}{c|}{\textbf{\textbf{Request responses}}} &
   \multirow{3}{*}{\begin{minipage}{0.7cm}\textbf{Cache hit ratio}\end{minipage}} \\
   \cline{2-5} \rule{0pt}{2ex}&
   \multicolumn{2}{c|}{\textit{\textbf{Cache}}} &
   \multirow{2}{*}{\textit{\textbf{Overlay}}} &
   \multirow{2}{*}{\textit{\textbf{Total}}} &
   \\
   \cline{2-3} \rule{0pt}{2ex}&
   \textit{Current} &
   \textit{Social} &
   &
   &
   \\
   \hline \rule{0pt}{2ex}\textit{Current} &
   3427562 &
   &
   197264 &
   3626229 &
   0.945 \\
   \hline \rule{0pt}{2ex}\textit{Social} &
   &
   5170354 &
   918365 &
   6090445 &
   0.849 \\
   \hline \rule{0pt}{2ex}\begin{minipage}{0.6cm}\textit{Current +Social}\end{minipage} &
   786123 &
   4606187 &
   44299 &
   5437792 &
   0.992 \\
   \hline
  \end{tabular}
 \end{minipage}%
 \hfill
 \begin{minipage}[b]{0.33\textwidth}
  \centering
  \tiny
  \caption{Cache Sizes and Responses/Item Compared}
  \label{table:eval_cs_leff_cache_size}
  \begin{tabular}{|p{0.6cm}|p{0.6cm}|p{0.6cm}|p{0.6cm}|p{0.6cm}|}
   \hline \rule{0pt}{2ex} \multirow{2}{*}{\begin{minipage}{0.7cm}\textbf{Cache setup}\end{minipage}} &
   \multicolumn{3}{c|}{\textbf{Items in cache}} &
   \multirow{2}{*}{\begin{minipage}{0.6cm}\textbf{Response / item}\end{minipage}} \\
   \cline{2-4} \rule{0pt}{2ex}&
   \textit{Current} &
   \textit{Social} &
   \textit{Total} &
   \\
   \hline \rule{0pt}{2ex}\textit{Current} &
   200674 &
   \textemdash &
   200674 &
   17.0802\\
   \hline \rule{0pt}{2ex}\textit{Social} &
   \textemdash &
   584968 &
   584968 &
   8.8386\\
   \hline \rule{0pt}{2ex}\textit{Current +Social} &
   50474 &
   889148 &
   939622 &
   5.7401\\
   \hline
  \end{tabular}
 \end{minipage}%
 
\end{table*}

From these values, we downsampled the experimental setup from 870 days a few days.
The downsampling of this information enables the evaluation of the caching approaches in smaller time frames without losing the usage characteristics gathered from the data set.
Our downsampling approach is shown in Equation~\ref{eq:sampling}
\begin{equation}
 \scriptsize
 \begin{aligned}
  sampledInterval(x) = \frac{dataSetExperimentTime}{newExperimentTime * x}
  \label{eq:sampling}
 \end{aligned}
\end{equation}
The
$sampledInterval(x)$ is the downsampled time period between an friend request or a wall post, where
$x$ represents the actual value calculated from the Facebook'09 data set.
The $dataSetExperimentTime$ is the whole time the original experiment has run, while
$newExperimentTime$ expresses the time on which the data is to be downsampled to.
This way, we can select a fixed number of 25 friends while ensuring the time span between new interactions and friend request remains relative to the analyzed Facebook'09 data set.
This helps attaining a similar utilization of user interactions in our test scenarios in comparison to the overall time period of the Facebook'09 experiment.

There are two test scenarios that are considered, each focusing on a different aspect.

\begin{itemize}
\item
 \textit{Comparing selection strategies}: In this paper, we propose three cache selection strategies. 
 We would like to find out which is the most suitable strategy considering all quality aspects.
 Therefore, for this test, the laboratory setup involved 64 LibreSocial instances on 8 distinct machines having similar hardware and operating system, Debian Linux.
 Each machine had 8 LibreSocial instances.
 We use a downsampled test scenario from the Facebook'09 data set which runs for 6 hours.
 For all tests only the social cache is enabled, but not the current caching mechanism.
 The test is for each of the three selection strategies, random, trend, and social score.
 For the parallel social update channels we choose the number of 15 simultaneously used channels, according to the Dunbar number, as well as a total MUC list size of 150 monitored users.
 The social bootstrapping functionality is enabled and the update interval is set to 50 seconds.
 A summary of the test setup can be found in Table~\ref{table:test_setup_parameters}.
\item
 \textit{Comparing caching approaches}: 
 We use the same parameters shown in Table~\ref{table:test_setup_parameters}.
 However, in this case, the test scenario runs for 2 days and thus the interaction and friend request intervals were downsampled to 2 days.
 Four tests are performed considering the following settings, \textit{caching completely disabled}, \textit{current caching only enabled} (LibreSocial before social caching), \textit{social caching only enabled} and both \textit{current and social caching enabled}.
\end{itemize}

\section{Results and Discussion}
We now consider the results obtained from the experiments conducted and analyze them for each test scenario in the following.

\subsection{Comparing selection strategies}
\label{subsec:comparing_selection_strategies}


The selection strategies are evaluated and the results are tabulated or graphed for comparison.
The performance of each of the strategies is shown in Table~\ref{table:eval_ss_perf}, the results for local efficiency and overall efficiency are shown in Fig.~\ref{fig:LocalEfficiency_Selection}.
We discuss them in detail.

\subsubsection{Performance}
From the values tabulated in Table~\ref{table:eval_ss_perf}, we calculate the cache hit ratio and compare the three selection strategies' performance.
The cache hit ratio ranges between 94.9\% and 92.4\%, which are very good values, as more than 92\% of the overlay lookups related to data requests have been saved. 
While the random selection strategy shows the highest cache ratio and the social score selection strategy has the lowest value, the values are very close.

\subsubsection{Local efficiency}


We now consider the local efficiency for the three selection strategies.
\begin{itemize}
\item
 \textit{Social cache size}: This is shown in Fig.~\ref{fig:eval_ss_leff_cache_size}.
 The social score selection strategy stores the smallest number of data objects within the cache structure, closely followed by the trend algorithm, with both algorithms maintaining roughly constant number of stored objects throughout the whole experiment.
 With the random selection strategy the social cache size keeps increasing.
 While the number of stored cache objects may not be directly related to the needed memory size to maintain them, we see that the random selection strategy needs a much higher amount of local resources to achieve 
 similar performance as trend and social score. 
 Thus the latter two approaches are much more efficient. 
\item
 \textit{Number of subscription messages}: Fig.~\ref{fig:eval_ss_leff_register} depicts the subscriptions during the experiments.
 The social score strategy has the least number of subscriptions hence fewer changes in the social cache participants.
 This correlates very well with the cache size results indicating high efficiency as it uses fewer local resources to achieve a high performance rate as evidenced by the cache hit ratio.
 While the trend strategy has more subscription messages than the social score strategy, it also uses the local resources well when we consider the social cache size.
 The results for the random selection strategy correlate with observations for the social cache size, having a large number of ever increasing subscriptions messages.
\item
\item
 \textit{Average size of MUC list}: This is depicted in Fig.~\ref{fig:eval_ss_leff_muc_size}.
 The social score and random strategies have almost similar values throughout.
 A low MUC list number can be seen as equivalent to a smaller amount of local resources used by the social cache and therefore a better \textit{local efficiency}.
 In comparison, the size of the MUC list for the trend strategy is constantly increasing contrary to expectations.
 This may need further investigation and a possible error may be due to the monitoring process rather than the actual MUC list structure.
\end{itemize}

Thus in view of the results gathered in analyzing the local efficiency, it can be inferred that overall, the social score strategy is the preferred choice over trend and random selection strategies as it attains a very high performance of 92.4\% cache hits against a drastically lower costs at the local node.

\subsubsection{Overlay efficiency} 
\begin{figure*}
 \centering 
 \scriptsize 
 \subfloat[Total social cache size]{\label{fig:eval_ss_leff_cache_size}\includegraphics[width = 0.26\linewidth]{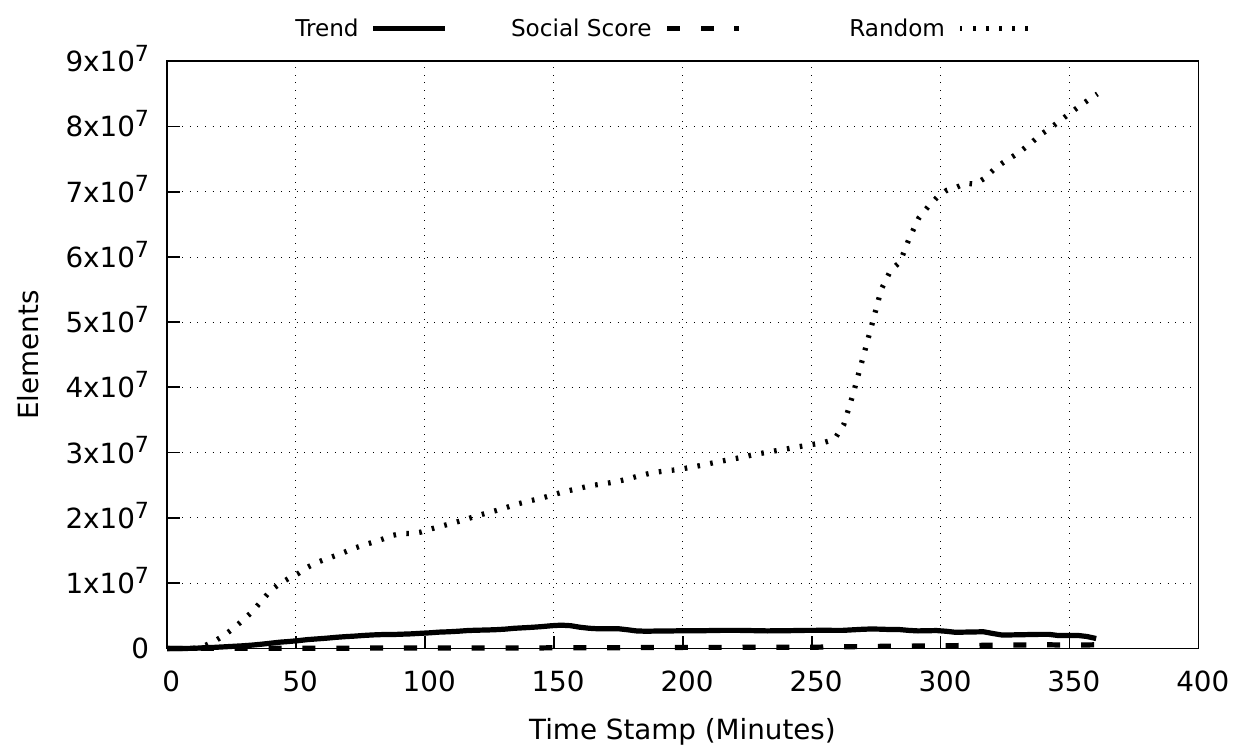}} 
 \subfloat[Social cache registration messages]{\label{fig:eval_ss_leff_register}\includegraphics[width = 0.25\linewidth]{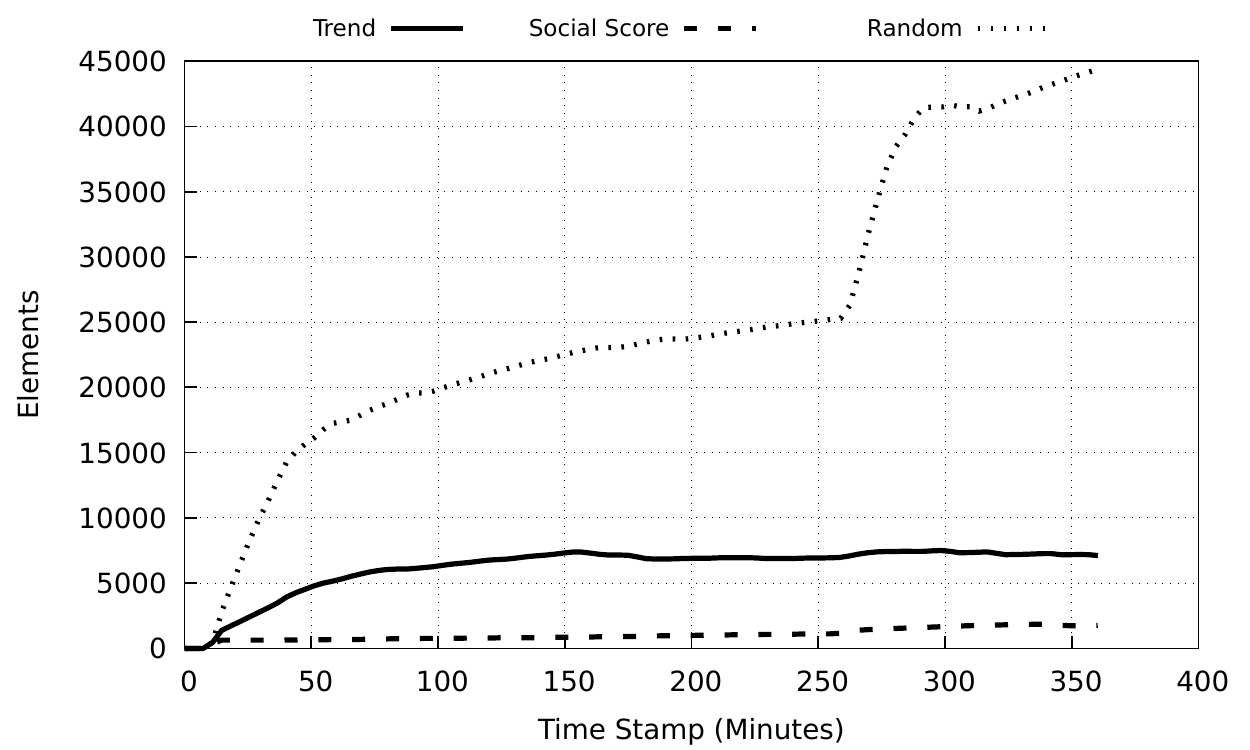}} 
 \subfloat[Average MUC list size/instance]{\label{fig:eval_ss_leff_muc_size}\includegraphics[width = 0.25\linewidth]{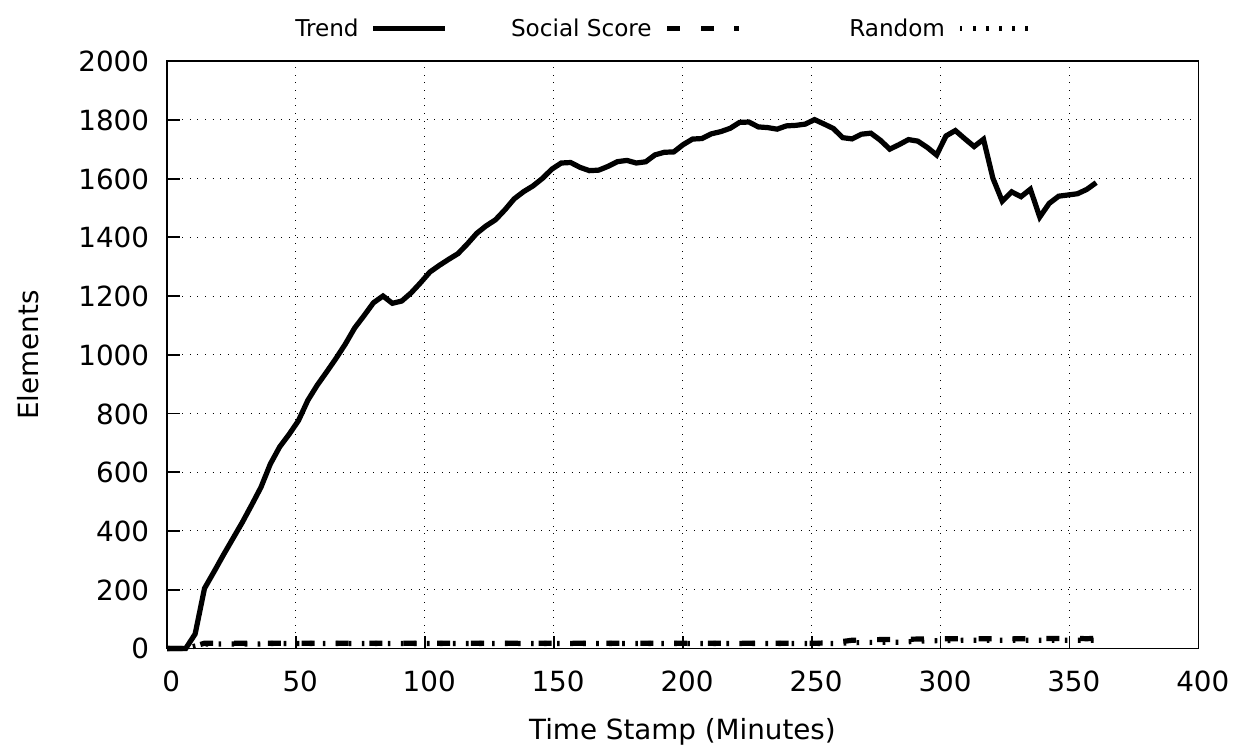}}\\
\vspace{-0.25cm}
 \subfloat[Data from the overlay]{\label{fig:eval_ss_oeff_pastry_sent}\includegraphics[width = 0.25\linewidth]{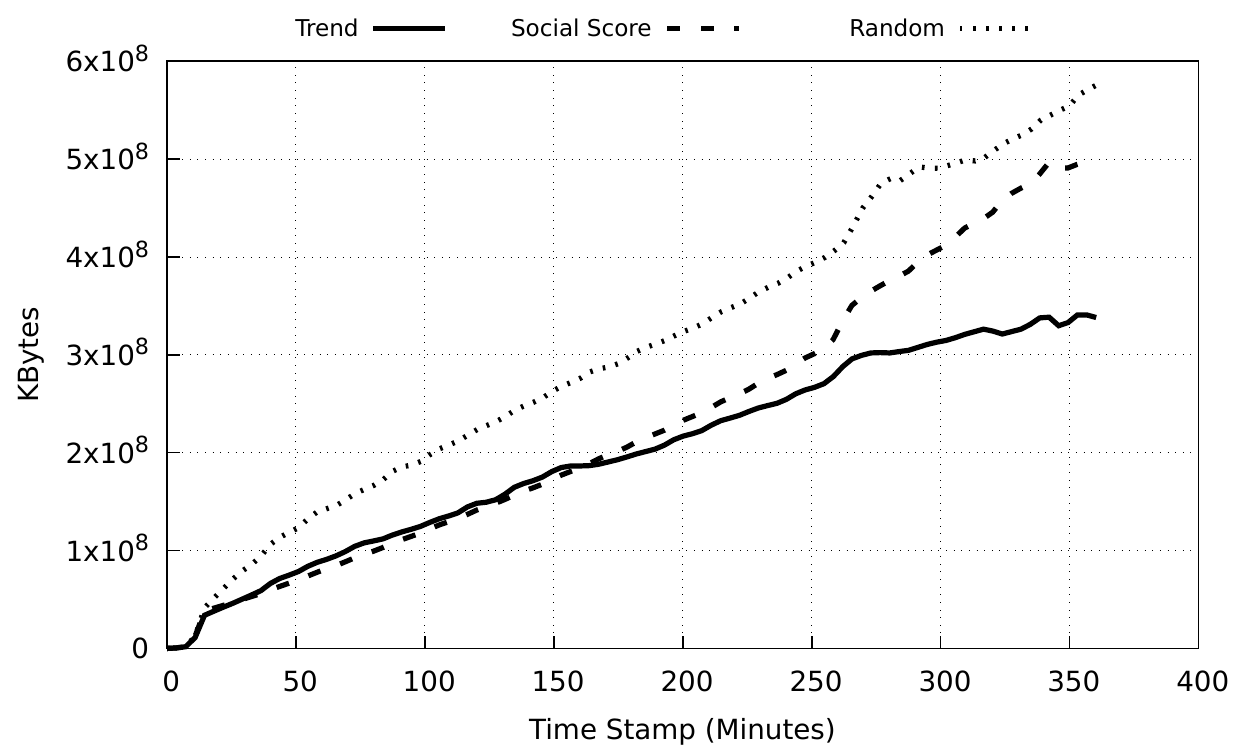}} 
 \subfloat[Average memory consumed/instance]{\label{fig:eval_ss_oef_memory_total}\includegraphics[width = 0.25\linewidth]{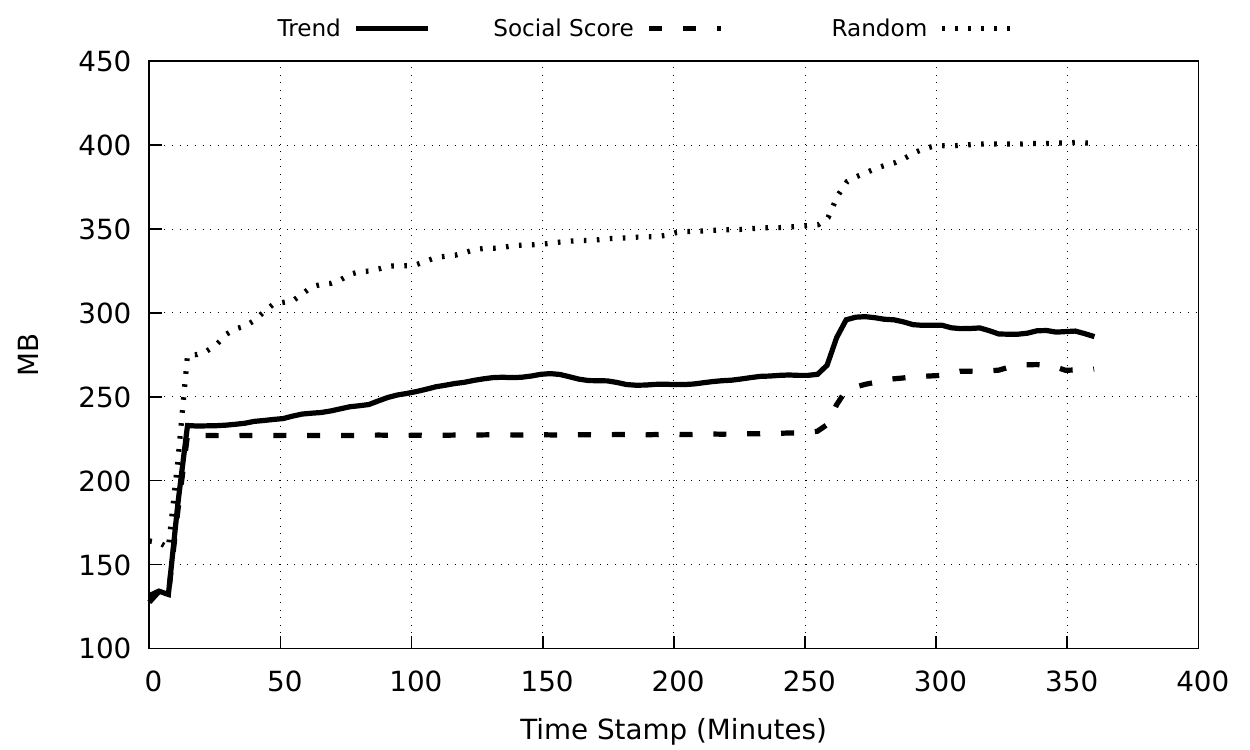}} 
 \subfloat[Messages at MessageDispatcher]{\label{fig:eval_ss_oeff_md_receive}\includegraphics[width = 0.25\linewidth]{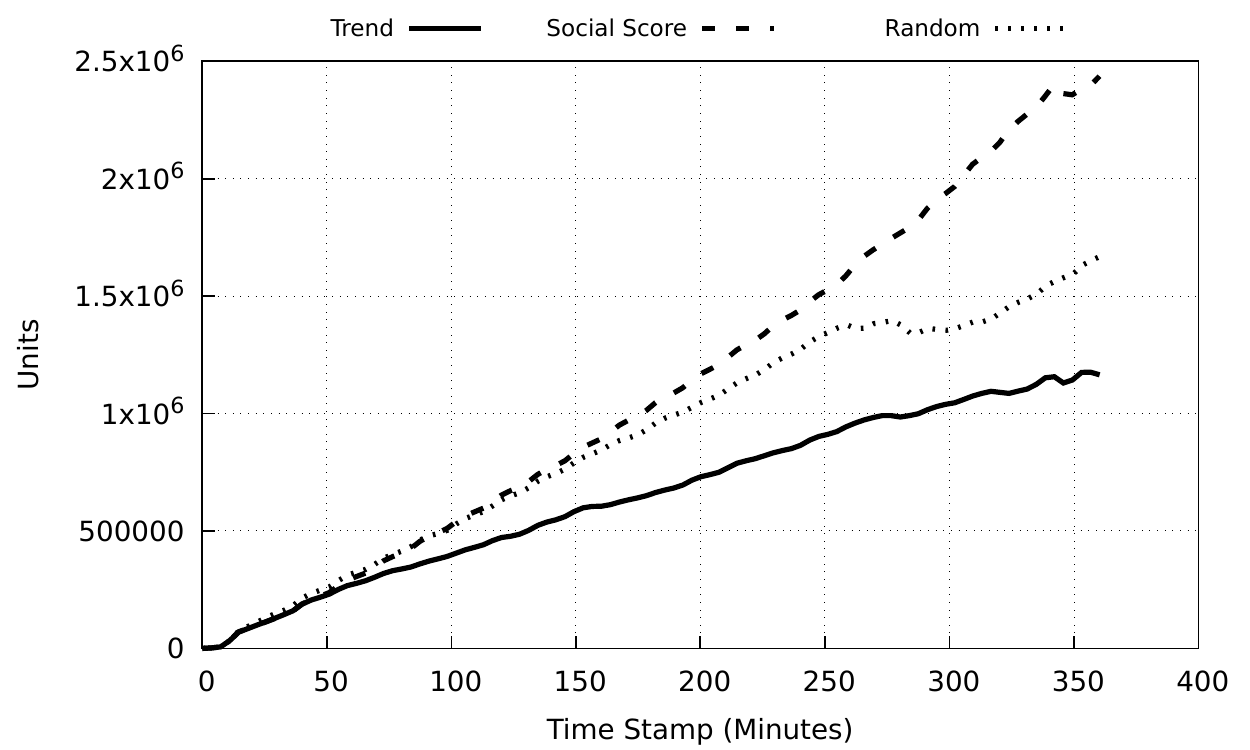}}\\
\vspace{-0.25cm}
 \subfloat[Total overlay data transferred]{\label{fig:eval_cs_oeff_pastry_receive}\includegraphics[width = 0.25\linewidth]{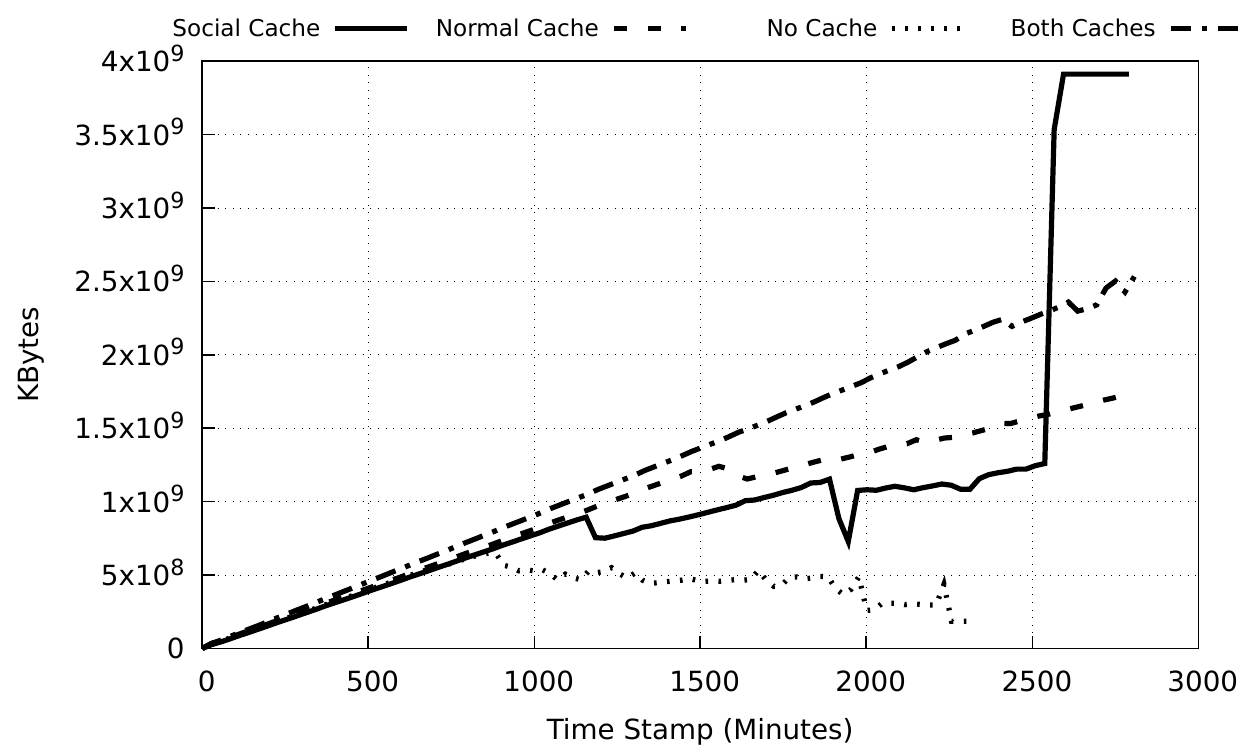}} 
 \subfloat[Average memory consumed/instance]{\label{fig:eval_cs_oef_memory_total}\includegraphics[width = 0.25\linewidth]{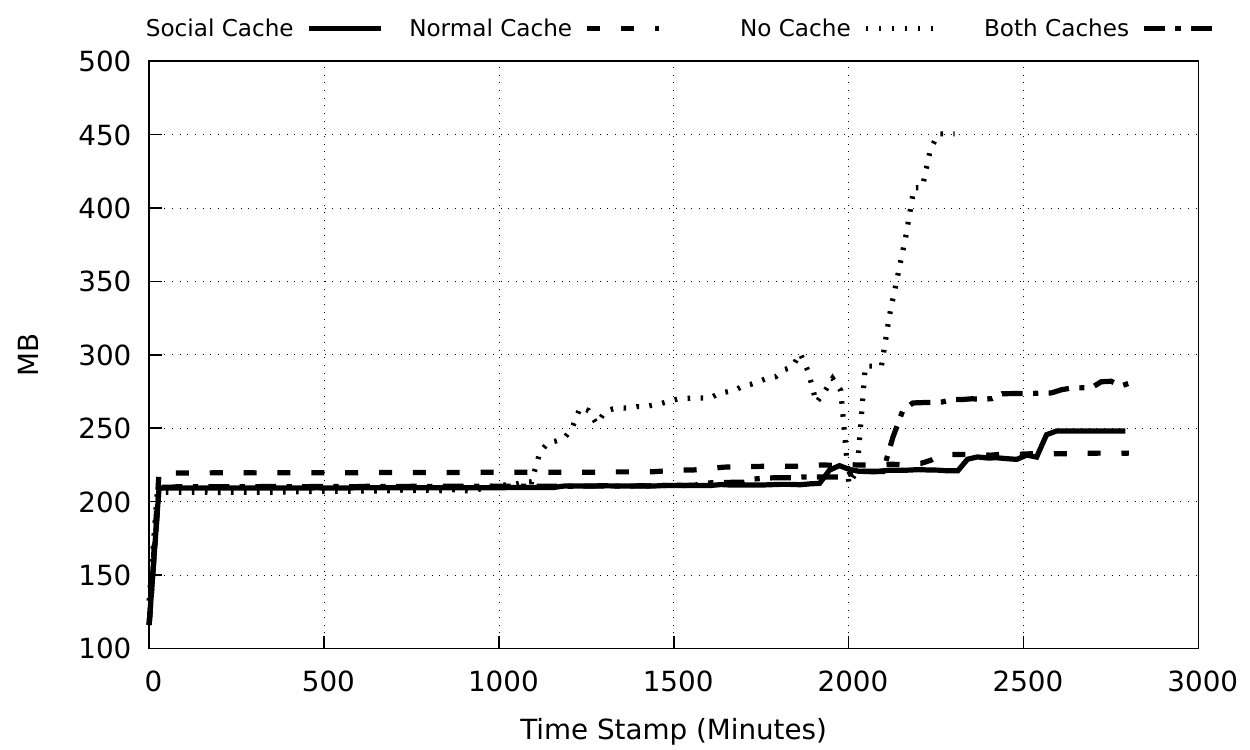}} 
 \subfloat[Messages at MessageDispatcher]{\label{fig:eval_cs_oeff_md_receive}\includegraphics[width = 0.25\linewidth]{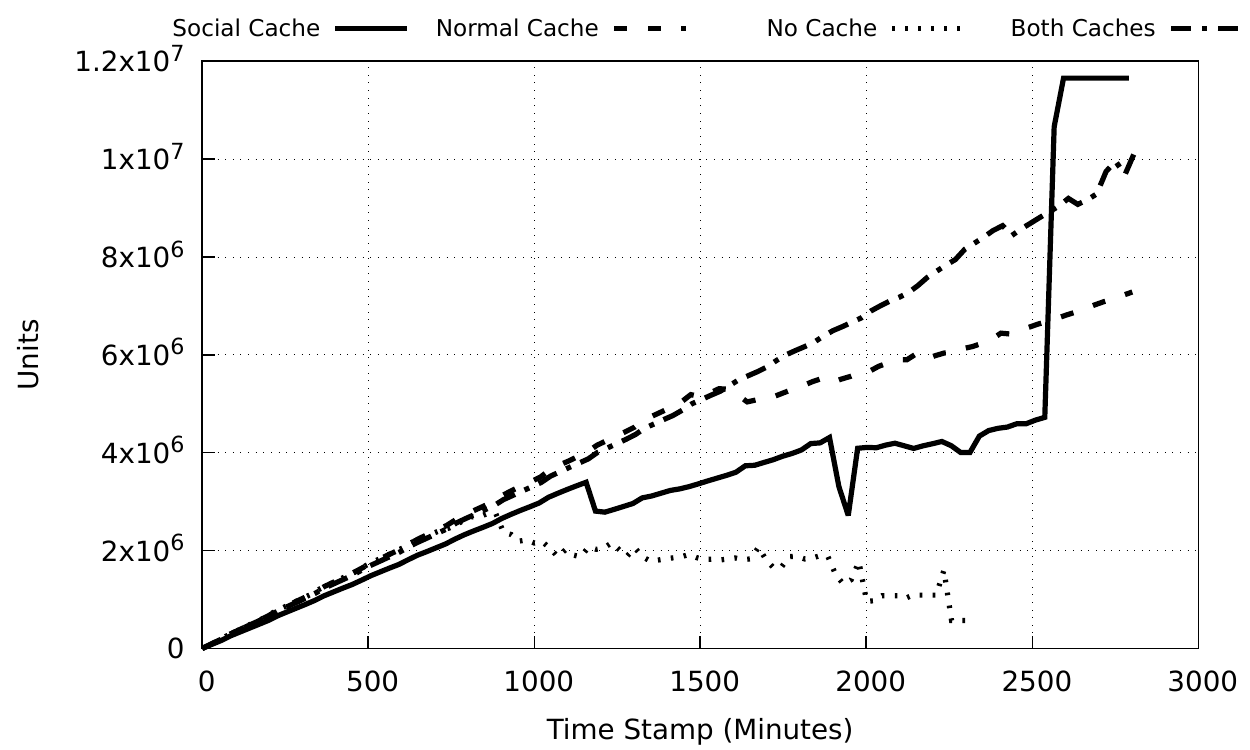}}
 \centering 
 \caption{Evaluation results of the local efficiency (resource utilization) and overall system efficiency}
 \label{fig:LocalEfficiency_Selection}
 \vspace{-0.25cm}
\end{figure*}

We analyze the overall efficiency of the selection strategies by considering the following overlay aspects.
\begin{itemize}
\item
 \textit{Overlay data load generated}: Fig.~\ref{fig:eval_ss_oeff_pastry_sent} shows the overall amount of data objects sent in the Pastry overlay for each selection strategy.
 Social score and trend selection strategies seem to have similar values of data sent until the second phase of friend requests is introduced to the system at about 257 minutes of the experiment.
 Then the trend strategy produces less data transactions than the social score strategy.
 The random selection strategy generates a higher amount of traffic due to overlay requests.
\item
 \textit{Average memory consumed per instance}: The results presented in Fig.~\ref{fig:eval_ss_oef_memory_total} are helpful in understanding the memory utilization for a single instance.
 For each instance, it can be seen that, depending on the selection strategy, average memory utilization is between 250 and 400 MB. Social score strategy has the least memory utilization followed very closely by the trend strategy, while the random selection strategy has the highest consumption of memory.
 In addition, it can be seen that the two times when friend requests occur have a considerable impact on the memory usage.
 The sudden sharp increases observed close to the end of the experiment is due to lookup request for unknown user leading to additional or change of subscriptions.
 The defined limit of parallel subscription channels therefore translates to a trade off between memory consumption and additional subscriptions for social updates.
\item
 \textit{Messages at the MessageDispatcher}: Focus here is on the messages produced at the overlay due to registration, updating, and bootstrapping processes of the social cache.
 The results can be seen in Fig.~\ref{fig:eval_ss_oeff_md_receive} in view of the MessageDispatcher messages.
 We see that the social score selection strategy generates the highest number of messages followed by the random strategy and finally the trend strategy.
 However, it is worth noting that more messages due to registration, updating and social bootstrapping is not a equivalent to a higher resource consumption, but can be an indicator for that.
 One reason for producing more messages may be the higher amount of total processed lookup request in the social score strategy, seen at the beginning of this evaluation.
\end{itemize}

In summary, the social score again shows the lowest costs in terms of data transfer and used memory, but produces more messages over the MessageDispatcher.
The trend selection strategy shows a similar pattern as the social score strategy albeit having slightly higher values for the data transfer and memory consumption but with the least messages observed a the MessageDispatcher.
The random selection strategy, may be considered to have poor efficiency when considering the data transferred and the memory consumed without consideration for the messages generated.

\subsection{Comparing caching approaches}
\label{subsec:comparing_caching_approaches}
From the results obtained in comparing the selection strategies, based on a comparison of the performance, local efficiency and overall efficiency, the most efficient selection strategy is the social score selection strategy.
Therefore, in the test scenarios that we conduct to compare the caching strategies, the social score strategy is used.
Before discussing the results of Figures \ref{fig:eval_cs_oeff_pastry_receive}-\ref{fig:eval_cs_oeff_md_receive}, it is important to note that the tests without any cache solution generally failed quarter way during the experimental period due to internal errors generated in Pastry.
Nevertheless, we include the results for comparison to show their influence on the data storage costs. 
We now discuss the results obtained from the cache comparison tests.

\subsubsection{Performance} Once again, the cache hit ratio is used to get the effectiveness of the cache mechanism when requests were sent by an instance.
The results are given in Table~\ref{table:eval_cs_perf} for three experiments, current (i.e. Pastry internal) cache only, the (new) social cache only, and both caches in combination.
With the social cache only, we reach a cache hit ratio of 84.9\%, which is nearly 10\% less than the current cache (94.5\%). 
Also the total amount of performed requests is much higher with the social cache only as compared to the current cache only, with 918,365 lookups  compared to 197,264 lookups.
However, when the two caching solutions are used in combination, a cache hit ratio of 99.2\% is observed which outperforms the current caching solution by nearly 5\%.
In this scenario, the current caching solution is accessed when the social cache does not answer the requests and therefore the current cache shows fewer cache hits in comparison to the social cache when in combination.
The social cache in this case responds to nearly 85\% of the cache answered requests.
It can thus be seen that the combination of both caching approaches leads to a more efficient a solution than having the two caching solutions separately.

\subsubsection{Local efficiency} We now look at the use of local resources for the three scenarios to set the performance in relation to utilization of the local system and therefore measure their efficiency.
Thus we focus on the number of stored items for each of the different caching structures.
In Table~\ref{table:eval_cs_leff_cache_size} the number of objects in the cache for each test scenarios as well as the ratio of requests per item is shown.
The number of items stored in the current cache alone as well as in the social cache solution alone are much fewer than when the two caches combined.
Comparing the current cache alone to the social cache alone, the current cache stores less than 50\% of the items in the social cache while initiating about twice as many cache responses per items (17 and 9 respectively).
Thus, the current cache appears to use the local resources more efficiently by achieving a higher cache hit ratio with fewer items in cache.
However, we have to contend with twice as many lookups.
The combination of both caches is accompanied by a drastic increase in the cached items (almost five times as much as the current caching only) but an equally significant drop in the number of cache responses per item (thrice as much as current cache only).
It is also observable, as expected, that there are much fewer items stored in the current cache as in the social cache.
This is explained by the hierarchy and arrangement of the caches.
The lower count of items in the current cache indicates a higher coverage of lookups responded to by the social cache.
Therefore it appears that in order to achieve a high performance as indicated by the hit ratio, the trade-off is a higher storage capability as indicated by the number of items stored by the combined solution local resource usage can be determined when it comes to use both caches at the same time.

\subsubsection{Overlay efficiency} Finally, we analyze the impact of the caching mechanisms in terms of generated workload at the overlay.
\begin{itemize}
\item
 \textit{Data transferred}: Fig.~\ref{fig:eval_cs_oeff_pastry_receive} shows the amount of data transferred and monitored in the Pastry layer of LibreSocial.
 The current caching solution causes more data transfer to occur at the overlay than the social caching solution.
 This may be due to the need to keep the data refreshed as a result of the validity time in cache for the current caching solution.
 There is a higher data transfer for the combined caching solution in comparison indicating higher bandwidth utilization.
 This is as both caches are operating simultaneously resulting in an addition on the communication payload and the needed lookup requests at the overlay.
\item
 \textit{Average memory consumed per instance}: This is illustrated in Fig.~\ref{fig:eval_cs_oef_memory_total}.
 The current cache and social cache when used separately consume similar amount of memory.
 For the combined caching approach, there is a similar trend until the second group of friend requests are made at about the 2200th minute of the experiment.
 Thereafter, the memory usage increases slightly in comparison to the other solutions.
\item
 \textit{Messages at MessageDispatcher}: The results are shown in Fig.~\ref{fig:eval_cs_oeff_md_receive}.
 The social cache alone generates the fewest messages to the overlay and the current cache alone. 
 It also has a higher number of messages, which indicates that not actively sharing content updates, as done by the current cache, results in the need for the system to exchange more messages with other system components.
 The combination of the two solutions caches has the highest amount of messages.
 We assume that the simultaneously use of both solutions tend result in a summing effect on resource consumption.
\end{itemize}

In summary based on the results of \textit{overlay efficiency}, the evaluation shows that the social cache has the least resource consumption as compared to the current caching approach.
However, the combined approach results in the highest overlay load.
By combining both efficiency qualities with the examined performances results, this evaluation in general shows that when the social cache is used alone, it has a lower cache hit ratio hence lower performance rating, while also consuming the most local resources in form of the managed caching structure and the corresponding need of memory to store it at a single instance.
However, at the overlay level the social cache shows slightly better results in the need of data and message transfer as well as memory usage.

The comparison on the current cache solution and the combination of the social cache and the current cache shows, that in terms of performance, this combination reaches by far the best cache hit ratio and can reduce the overlay request to 0.8\% of the total used lookups in the Information Cache.
This performance advantages comes with the trade off of a higher load and resources consumption at the local level as well as in the overlay.
It is also important to notice, that the use of the social cache ensures the consistency of the cached and used information through the social update mechanism.
This consideration can not be guaranteed in the current cache mechanism.

\section{Conclusion}
The use of a social caching mechanism presents several advantages but comes with several trade offs that must be put into consideration.
The use of the social interactions to actively select where to replicate data in combination with an active dissemination strategy for information diffusion makes for a good solution to ensure that the data is always up to date.
We present in this paper three selection strategies, random, trend and social score, and show that the social score strategy is the best solution for active dissemination of content.
Using LibreSocial as our testing platform, we have implemented a social caching mechanism on top of the existing request-reply caching mechanism and compared the performance and efficiency of the two solutions as well as the combination of the two.
The combination of the two caching solutions gave the best performance as recorded by the cache hit ratio value but with a compromise on the local as well as overall resources.
For the user it seems to be of higher value to have a very low delay in the user interface and having instant presentation of social network elements he desires. 
For that it seems reasonable to apply such a combined social and conventional caching approach leading to a higher user satisfaction. 
The corresponding higher load in the networking and storage utilization is of lower relevance, as long as other services run on the computer are not disturbed. 
Thus, our proposed approach presents a clear benefit to the user. 

\bibliographystyle{IEEEtran2}
\balance
\bibliography{references}
\end{document}